\documentclass[twocolumn,showpacs,amsmath,amssymb,preprintnumbers]{revtex4}

\usepackage{graphicx}
\usepackage{dcolumn}
\usepackage{bm}

\newcommand{\beq}{\begin{eqnarray}}
\newcommand{\beqa}{\begin{eqnarray*}}
\newcommand{\eeq}{\end{eqnarray}}
\newcommand{\eeqa}{\end{eqnarray*}}
\newcommand{\tr}{{\rm Tr}}

\newcommand{\nn}{\nonumber}

\newcommand{\del}{\partial}
\newcommand{\dif}[1]{\frac{\del}{\del #1}}
\newcommand{\difto}[1]{\frac{\del^2}{\del #1}}
\newcommand{\ba}{\begin{array}}
\newcommand{\ea}{\end{array}}
\newcommand{\z}[1]{{\cal Z}^{\beta=#1}}
\newcommand{\mb}[1]{\mbox{\bf{#1}}}
\newcommand{\M}{{\cal M}}

\begin{document}

\title{Finite Volume Chiral Partition Functions and the Replica
Method}

\author{Jesper Levinsen}
\email{levi@nbi.dk}
\affiliation{The Niels Bohr Institute\\ Blegdamsvej 17, DK-2100
Copenhagen {\O}\\ Denmark}

\date{\today}

\begin{abstract}
In the framework of chiral perturbation theory we demonstrate the
equivalence of the supersymmetric and the replica methods in the
symmetry breaking classes of Dyson indices $\beta=1$ and $\beta=4$.
Schwinger-Dyson equations are used to derive a universal differential
equation for the finite volume partition function in sectors of fixed
topological charge, $\nu$. All dependence on the symmetry breaking
class enters through the Dyson index $\beta$. We utilize this
differential equation to obtain Virasoro constraints in the small mass
expansion for all $\beta$ and in the large mass expansion for
$\beta=2$ with arbitrary $\nu$. Using quenched chiral perturbation
theory we calculate the first finite volume correction to the chiral
condensate demonstrating how, for all $\beta$ there exists a region
in which the two expansion schemes of quenched finite volume chiral
perturbation theory overlap.
\end{abstract}

\pacs{11.30.Rd, 12.39.Fe, 12.38.Lg}

\maketitle


\section{Introduction}
In the low-energy and chiral limit of QCD, chiral symmetry is
spontaneously broken. The physics is dominated by the pseudo-Goldstone
bosons, and the corresponding effective field theory is known as
chiral perturbation theory. This theory is describable in terms of a
chiral Lagrangian. However, the physics in this low-energy regime
depends upon the exact way in which the chiral symmetry is
spontaneously broken. There are believed to be three ways in which
spontaneous chiral symmetry breaking can happen \cite{P80}. These
depend upon the representation of the fermions in the following way:
\begin{itemize}
\item{The representation of the fermions is pseudo-real. The expected
  symmetry breaking pattern is $SU(2N_f)\to Sp(2N_f)$.}
\item{The representation is complex. In this case we expect the
  symmetry breaking pattern to be $SU_L(N_f)\times SU_R(N_f)\to
  SU(N_f)$. An example is fermions in the fundamental representation
  with the number of colours, $N_c\geq 3$. Thus ordinary QCD falls
  into this class.}
\item{The fermions are in a real representation. The expected
  symmetry breaking pattern is then $SU(N_f)\to SO(N_f)$.}
\end{itemize}
These symmetry breaking patterns are nowadays labeled by the
Dyson-indices, $\beta=1$, $\beta=2$, and $\beta=4$, respectively, due
to a connection to Random Matrix theory \cite{V94}. For a symmetry
breaking pattern $G\to H$, the fields live on the coset $G/H$. The
usual perturbation scheme for chiral perturbation theory is for large
volumes and is an expansion in terms of the momenta of the Goldstone
modes. This perturbative scheme will be applied in most of this paper.

A systematic approach to calculations in (partially) quenched chiral
perturbation theory is known as the supersymmetric method
\cite{M87,BG92,S92}. In the supersymmetric formulation $k$ ``valence''
quark species are introduced with $k$ ghost quarks of opposite
(bosonic) statistics in addition to the $N_f$ ``sea'' quarks. In this
way the effective partition function is extended, becoming a generator
of $n$-point functions with the additional quarks acting as source
terms. The terminology ``supersymmetric method'' reflects the fact
that the chiral flavor symmetry group is extended to a super Lie
group, while space-time supersymmetry is not intrinsic to the method.

The replica method, which shall be applied in this paper, has turned
out to be an alternative to the supersymmetric method. In this method
one adds $N_v$ fermionic valence quarks taking $N_v$ to zero at the
end of the calculations. Obviously, if we let $N_v\to 0$ we recover
the original partition function. However, as in the supersymmetric
method this extension of the partition function makes it a generating
functional for $n$-point functions with the sources being the valence
quarks. It can be convenient to add $k$ sets of $N_v$ valence quarks
with masses $m_{v_1},\dots,m_{v_k}$, but for the applications in this
paper we need only include one set of $N_v$ valence quarks of mass
$m_v$. We assume the symmetry breaking patterns in this extended
theory to be the usual symmetry breaking patterns. Thus for
intermediate calculations in this paper, in the above-mentioned
symmetry breaking patterns we will simply replace $N_f$ by
$N_f+N_v$. Since the limit $N_v\to 0$ formally requires an analytic
continuation of the Lie groups to non-integer dimensions the validity
of the method is not obvious, but it can be shown that the required
analytic continuation can be performed in series expansions
\cite{DS00,DS00A,DV01}. Recently there have been some developments
into non-perturbative results in this framework \cite{KSV02}, but
these are out of the scope of this paper. In both this framework and
the supersymmetric one, the fully quenched limit corresponds to
vanishing $N_f$ while the theory is partially quenched for $N_f$
non-vanishing.

The equivalence of these methods have been demonstrated for $\beta=2$
and thus for ordinary QCD \cite{DS00} by explicitly showing the
equivalence of the Feynman rules at the one-loop level. In the first
part of this paper we will extend this analysis to the other classes
of chiral symmetry breaking, showing the equivalence only explicitly
at the one-loop level but noting that the analysis is easily carried
over to the higher loop level. To illustrate the equivalence we will
also calculate the partially quenched chiral condensate at one-loop
order, illustrating how the statistics signs of the supersymmetric
formulation are reproduced in the replica formulation by the counting
of possible quark-loops.

In the second part of this paper we present one of the main results of
this paper, a universal second order differential equation to
determine the finite volume chiral partition function in all classes
of chiral symmetry breaking and for arbitrary topological charge,
$\nu$. This differential equation is written in terms of the fermion
masses and all dependence on the pattern of symmetry breaking is
through the Dyson index, $\beta$. The form of the differential
equation in the specific case of $\beta=2$ with zero topological
charge is already known \cite{GN92}, but the generalization of this
paper has not been found previously. The method is to develop
Schwinger-Dyson equations for the effective low energy, or finite
volume, partition function and next to realize how these can be written in
terms of the fermion masses.  For $\beta=2$ with vanishing topological
charge the effective partition function has been recognized
\cite{MMS96} as a group integral of Kontsevich type with potential
${\cal V}(X)=1/X$ and this model, also known as the one-link integral,
has been intensively studied in the literature, see, e.g., refs.
\cite{BN81, D00, DS00A, BRT81}.

Having found a governing differential equation we proceed to solve
this equation. It should be realized that there are two relevant
expansions, the small-mass and the large-mass expansions.  Using the
invariance properties of the effective finite volume partition
function it is possible to find Virasoro constraints.  For the small
mass phase this was first done in ref. \cite{MMS96} in the simplest
version, namely $\beta=2$ with vanishing topological charge, by
recognizing the partition function as being a unitary integral of
Kontsevich type.  More recently, Virasoro constraints in the small
mass phase for general $\beta$ and topological charge have been
calculated in ref. \cite{DV01} by using a method much similar to the
one presented here. We exactly reproduce the Virasoro constraints of
ref. \cite{DV01}. The large mass expansion is more involved as it is
not possible to do a simple perturbative expansion, instead the large
mass expansion is an asymptotic, or saddle-point, expansion
\cite{GN92,MMS96}. The saddle-point approximation corresponds to the
classical limit and is an expansion in powers of $1/N_f$. In the large
mass limit Virasoro constraints have been previously determined in the
case of $\beta=2$ \cite{GN92}. We find Virasoro constraints in the
general case of $\beta=2$, where the inclusion of a non-vanishing
topological charge is thus a new result. At next to leading order in
$1/N_f$ the saddle-point approximation turns out to be a simple
correction to the leading order term in addition to an (apparently)
infinite expansion in the inverse fermion masses. However, this last
expansion is proportional to $1-2/\beta$ and thus it vanishes if
$\beta=2$. Precisely this property allows us to calculate large-mass
Virasoro constraints in the $\beta=2$ case with arbitrary topological
charge. Although we do not determine Virasoro constraints in the other
classes of symmetry breaking we find that it is nevertheless possible
to extract useful information from the saddle-point approximation; in
the case of equal fermion masses the differential equation for the
partition function is still tractable. We solve the governing
differential equation for equal fermion masses and as an application
we find the lowest order corrections to the fully quenched chiral
condensate in sectors of topological charge, a result which will be
very useful in the last part of this paper.

As Gasser and Leutwyler \cite{GL87} were the first to point out, there
are two finite volume regimes to consider. The first is the large, but
finite, volume which can be considered a small perturbation to the
infinite volume theory. This is the volume considered in the first
parts of this paper and to which the usual chiral perturbation theory
in terms of a momentum expansion of the chiral Lagrangian applies.
The last part of the paper concerns quenched chiral perturbation
theory in volumes much smaller than the correlation lengths of the
Goldstone modes. Although this may seem ill-defined at first it is
still possible by means of numerical simulations to extract physical,
infinite volume quantities from this theory. In this volume-regime the
usual momentum expansion breaks down due to the presence of
zero-momentum modes \cite{N88}. The different expansion scheme
required in this regime, known as the $\epsilon$-expansion
\cite{GL87}, treats this problem by means of a collective field
technique, collecting the zero-momentum modes in one field and the
non-zero modes in another. The non-zero modes are still
perturbative. Difficulties in this scheme arise from the fact that the
integration over the zero-momentum modes has to be done exactly.

In ref. \cite{D01} it was shown that in the case of $\beta=2$ it is
possible to determine a region in which the two perturbative expansion
schemes overlap also in the quenched limit. We extend this analysis
to the other classes of chiral symmetry breaking, seeing how also in
these cases such a region exists and demonstrating how the quenched
chiral condensates of the two regimes exactly match in the
above-mentioned region.

This paper is organized as follows. In Section \ref{sec:equiv} we
consider (partially) quenched chiral perturbation theory. The
equivalence of the supersymmetric and the replica method in the
alternative ($\beta=1,4$) classes of spontaneous chiral symmetry
breaking is explicitly demonstrated at the one-loop level. Since
equivalence was demonstrated for ordinary QCD ($\beta=2$) in ref.
\cite{DS00}, these two approaches to calculations in (partially)
quenched chiral perturbation theory are equivalent in all classes of
spontaneous chiral symmetry breaking.

Section \ref{vcsec} concerns finite volume chiral perturbation
theory.  Schwinger-Dyson equations are derived for the finite volume
chiral partition functions and utilized to obtain Virasoro constraints
in the limit of small quark masses and also (for $\beta=2$) in the
limit of large quark masses. For general $\beta$ we find the next to
leading order term in a mass expansion of the partition
function. Although the calculations of this section are not considered
to be in a quenched setting, the relevance of these results in such a
setting is demonstrated by calculating the first correction in a mass
expansion to the quenched chiral condensate.

We return to considering quenched chiral perturbation theory in section
\ref{sec:evsp}. Here it is demonstrated how a volume-regime exists in
which the two perturbation schemes of finite volume chiral
perturbation theory match in the quenched approximation.


\section{The equivalence of the supersymmetric and the replica methods
\label{sec:equiv}}
It has been shown that an alternative approach towards calculations in
partially quenched chiral perturbation theory is offered by the
replica method. The equivalence between the supersymmetric method and
the replica method has been demonstrated for $\beta=2$ in ref.
\cite{DS00}. In chiral perturbation theory the natural way of
demonstrating the equivalence between the methods is demonstrating how
the generating functional in the replica formulation results in
Feynman rules equivalent to those obtained from the generating
functional in the supersymmetric formulation. In this section we will
extend this equivalence to the other classes of chiral symmetry
breaking.  As was the case in ref. \cite{DS00} we will only
demonstrate this equivalence at the one-loop level and thus the ${\cal
O}(p^2)$ expansion of the Lagrangian density is sufficient with the
${\cal O}(p^4)$ terms acting as counter terms. To demonstrate the
equivalence, we compare our results with those obtained in ref.
\cite{TV99} by means of the supersymmetric method.  The extension to
the ${\cal O}(p^4)$ Lagrangian is not very difficult, as it only
affects the vertices, and not the propagator structure.

In the framework of chiral perturbation theory,
the effective partition function is given by
\beq
{\cal Z} = \int_{U\in G/H}dU ~e^{-\int d^4x{\cal L_{\rm eff}}}
\eeq
with the usual effective chiral Lagrangian density to second order in
momenta
\beq
{\cal L}_{\mbox{eff}} & = &
\frac{F_\beta^2}{4}\tr(\partial_\mu U^\dag\partial_\mu U) -
\frac{\Sigma_0}{2}\tr({\cal M}^{(\beta)}(U+U^\dag)) \nn \\
& & + \frac{m_0^2}{2N_c}\Phi_0^2+
\frac{\alpha}{2N_c}\partial_\mu\Phi_0\partial_\mu\Phi_0.
\label{lagrangian}
\eeq

The Goldstone fields $U$ are parameterized as
\beq
U=\exp(i\sqrt{2}\Phi/F_\beta).
\eeq
The last two terms in (\ref{lagrangian}) contain the flavor singlet
field $\Phi_0 \equiv \tr \Phi$. In the supersymmetric formulation this
term is an invariant of the reduced symmetry group and it has to be
included in the replica Lagrangian as well, since otherwise the
quenched replica limit would not exist \footnote{The parameter $m_0$
allows for a smooth interpolation between the coset $G/H$ and
$U(1)\times G/H$. There is no fully quenched replica limit of the
three cosets mentioned in the introduction due to the determinant
constraint.}. For the time being we will ignore the term proportional
to $\alpha$ and reinstate it at the end of this section by
substituting $m_0^2$ with $m_0^2+\alpha p^2$.

The pion decay constant $F_\beta$ is chosen so that the fields satisfy the
usual Gell-Mann--Oakes--Renner relation for the masses of the mesons
\beq
M_{ij}^2 & = & (m_i+m_j)\frac{\Sigma_0}{F^2}.
\label{gmor}
\eeq
With our choice of normalization of the fields it turns out that
\beq
F_2 = F_4 = F\\
F_1 = F/\sqrt{2}.
\eeq
Notice that the fact that our choices of normalization of the fields
and of $F_\beta$ differ from those of ref. \cite{TV99} does not stem from
any peculiarities of the replica method. Rather, we find these choices
more natural, since they ensure that all complex fields are normalized
in the same way.

The infinite volume chiral condensate is denoted by $\Sigma_0$ and
in the replica formulation the partially quenched chiral condensate of the
additional fermionic copies is, to one-loop order, given by
\beq
\Sigma(m_v,\{m\}) & = &
\frac{1}{V}\lim_{N_v\to 0}\frac{1}{N_v}\dif{m_v}\ln{\cal Z} \nn \\
& =& \lim_{N_v\to 0}\frac{1}{N_v}\left<\del_{m_v}(-{\cal L})\right>.
\label{cc}
\eeq


\subsection{$\beta=1$}
The expected symmetry breaking of fermions in a pseudo-real representation
is
$SU(2N_f+2N_v)\rightarrow Sp(2N_f+2N_v)$ with Dyson index $\beta=1$.

Following ref. \cite{P80} we choose an explicit representation of the
field
$\Phi$ as
\beq
\Phi = \frac{1}{\sqrt{2}}\left(\ba{cc} \phi & \psi \\
\psi^\dagger & \phi^T\ea\right).
\label{fieldbeta1}
\eeq
The normalization is taken for convenience. In this representation the
field $\phi$ is Hermitian while the field $\psi$ is complex and
anti-symmetric. Both fields are written in terms of
$(N_f+N_v)\times(N_f+N_v)$
matrices.

The mass matrix can be chosen diagonal with
\beq
{\cal M}^{(1)} & = \frac{1}{2} ~ \mbox{diag}(&m_1,\dots, m_{N_f},
\overbrace{m_v, \dots, m_v}^{N_v}, \nn \\
&& m_1,\dots, m_{N_f}, \overbrace{m_v, \dots, m_v}^{N_v} \:),
\label{massbeta1}
\eeq
where this normalization ensures that the Gell-Mann--Oakes--Renner
relation (\ref{gmor}) is satisfied.

To calculate the chiral condensate to one-loop order we evaluate the
Lagrangian to second order in the fields.  Only the term containing
$m_0^2$ mixes states and this only applies to the diagonal
fields. This makes it easy to read off the propagators for the
``off-diagonal'' mesons. These are seen to be
\beq
D_{ij}=\frac{1}{p^2+M_{ij}^2}.
\label{delta}
\eeq
The diagonal mesons also inherit this part, so that their propagator
becomes
\beq
G_{ij}^{-1} = \delta_{ij}(p^2+M_{ii}^2)+2m_0^2/N_c.
\eeq
This expression can be inverted by noting that for general matrices
$A, B$, with $A_{ij}=a$ constant, and $B_{ij}=\delta_{ij}b_{ii}$
(no sum over $i$) diagonal,
\beq
(A+B)^{-1} = B^{-1}-B^{-1}AB^{-1}\frac{1}{1+a\sum_{i=1}^{\cal
N}b_{ii}^{-1}}.
\label{tvtrick}
\eeq
Thus we find that
\beq
G_{ij} = \frac{\delta_{ij}}{p^2+M^2_{ij}}-
\frac{2m_0^2/N_c}{(p^2+M_{ii}^2)
(p^2+M_{jj}^2){\cal F}^{\beta=1}(p^2)},
\eeq
where
\beq
{\cal F}^{\beta=1} \! & = & \!
1+\frac{2m_0^2}{N_c}\sum_{k=1}^{N_f+N_v}(p^2+M_{kk}^2)^{-1} \nn \\
& = & \! 1\!+\!\frac{2m_0^2}{N_c}\left( \frac{N_v}{p^2+M_{vv}^2}
+ \sum_{k=1}^{N_f}\frac{1}{p^2+M_{kk}^2}\right)\!,
\eeq
since the masses of the mesons in the replica set are identically
$m_v$. These expressions for the propagators can be compared with the
corresponding expressions from the supersymmetric method obtained in
\cite{TV99}.  For $N_v\to 0$ the expressions are equivalent, the
difference arising from our alternative normalization conventions,
only.

We now proceed to calculate the chiral condensate to one-loop order.
It is given by (\ref{cc}) and can thus be calculated directly from
the second order expansion of the Lagrangian. We find that
\beq
&&\hspace{-1cm}
\frac{\Sigma(m_v,\{m\})}{\Sigma_0} = \frac{1}{V\Sigma_0}\lim_{N_v\to
0}\frac{1}{N_v}\dif{m_v}\ln{\cal Z} \nn \\ & = & \lim_{N_v\to
0}\frac{1}{N_v}
\left[N_v-\frac{2}{F^2}\left(N_v\sum_{f=1}^{N_f}\Delta_{vs}
\right.\right. \nn \\ && \left.\left. + N_v(N_v-1)
\Delta_{vv}+\frac{N_v}{2V}\sum_p G_{vv}(p^2)
\vphantom{\sum_{f=1}^{N_f}}\right)\right] \nn\\
& = & 1-\frac{1}{F^2}
\left(2N_f\Delta_{vs}-2\Delta_{vv}+\frac{1}{V}\sum_p G_{vv}\right),
\eeq
where the pion propagator is given by
\beq
\Delta_{ij} & \equiv & \frac{1}{V}\sum_p\frac{1}{p^2+M^2_{ij}} =
\frac{1}{V}\sum_p D_{ij}(p^2),
\label{pprop}
\eeq
and where we in the last step for simplicity have put the sea quark
masses equal. This result is identical to the result obtained from
the supersymmetric formulation \cite{TV99}.


\subsection{$\beta=4$}
For fermions in a real representation the expected symmetry breaking class
is $SU(N_f+N_v)\rightarrow SO(N_f+N_v)$ corresponding to the Dyson index
$\beta=4$. In this case an explicit representation of the Goldstone bosons
is \cite{TV99}
\beq
\Phi=\left(\begin{array}{cccc}
A_{11} & \frac{1}{\sqrt{2}}A_{12} & \cdots &
\frac{1}{\sqrt{2}}A_{1,N_f+N_v} \\
\frac{1}{\sqrt{2}}A_{12} & \ddots& & \vdots\\
\vdots& & \ddots & \vdots \\
\frac{1}{\sqrt{2}}A_{1,N_f+N_v}& \cdots& \cdots & A_{N_f+N_v,N_f+N_v}
\end{array}\right)\!\!.
\label{fieldbeta4}
\eeq
$\Phi$ is a real, symmetric matrix \cite{P80}.
Thus the normalization of the off-diagonal mesons
differs from that of the diagonal mesons to ensure
correctly normalized kinetic terms in
the Lagrangian. The mass matrix is diagonal with
\beq
{\cal M}^{(4)}=\mbox{diag}(m_1,\dots, m_{N_f},
\overbrace{m_v, \dots, m_v}^{N_v}).
\label{massbeta4}
\eeq

Following the steps above we evaluate the Lagrangian (\ref{lagrangian})
to second order in the fields to obtain the
one-loop correction to the chiral condensate.
As for $\beta=1$ the propagators of the off-diagonal mesons are easy to
read
off. Again we find
\beq
D_{ij}=\frac{1}{p^2+M_{ij}^2}.
\eeq
The propagators of the diagonal mesons are
\beq
G_{ij}^{-1}=\delta_{ij}(p^2+M_{ii}^2)+m_0^2/N_c
\eeq
which, inverted by the use of eq. (\ref{tvtrick}), yields
\beq
G_{ij} = \frac{\delta_{ij}}{p^2+M_{ij}^2}
-\frac{m_0^2/N_c}{(p^2+M_{ii}^2)(p^2+M_{jj}^2)
{\cal F}^{\beta=4}(p^2)},
\eeq
with
\beq
{\cal F}^{\beta=4}
=
1+\frac{m_0^2}{N_c}\left(\frac{N_v}{p^2+M_{vv}^2}
+\sum_{k=1}^{N_f}\frac{1}{p^2+M^2_{kk}}\right).
\eeq
In the limit $N_v\to 0$ this result also agrees with the result
obtained in ref. \cite{TV99} from the supersymmetric formulation.

The calculation of the partially quenched chiral condensate proceeds
as before. From (\ref{cc}) we find
\beq
\frac{\Sigma(m_v,\{m\})}{\Sigma_0}
& = & 1-\frac{1}{F^2}\left(\vphantom{\sum_p}\frac{1}{2}N_f\Delta_{vs}
-\frac{1}{2}\Delta_{vv} \right. \nn \\
&& \hspace{15mm} \left.+\frac{1}{V}\sum_pG_{vv}(p^2)\right),
\eeq
with the pion propagators given as in eq. (\ref{pprop}) and the sea quark
masses equal.
Again, this result exactly matches the result from the supersymmetric
method
\cite{TV99}.


\subsection{The chiral condensate in all classes of chiral symmetry
breaking}
Including the chiral condensate in theories with fermions in
a complex representation ($\beta=2$) \cite{GL98,OTV99},
allows us to write, for general $\beta$,
\beq
\frac{\Sigma^\beta(m_v,\{m\})}{\Sigma_0}
& = & 1-\frac{1}{F^2}\left(\vphantom{\sum_p}\frac{2}{\beta}N_f\Delta_{vs}
-\frac{2}{\beta}\Delta_{vv} \right. \nn \\
&& \hspace{15mm} \left.+\frac{1}{V}\sum_pG_{vv}(p^2)\right),
\label{ccgenbeta}
\eeq
with the propagator for the neutral mesons
\beq
G_{ij} = \frac{\delta_{ij}}{p^2+M_{ij}^2}
-\frac{(1+\delta_{\beta,1})(m_0^2+\alpha
p^2)/N_c}{(p^2+M_{ii}^2)(p^2+M_{jj}^2)
{\cal F}^{\beta}(p^2)}.
\label{diagprop}
\eeq
For general $\beta$
\beq
{\cal F}^\beta & = &
1+(1+\delta_{\beta,1})
\frac{m_0^2+\alpha p^2}{N_c}\left(\frac{N_v}{p^2+M_{vv}^2} \nn
\vphantom{\sum_{k=1}^{N_f}}\right. \\
&&\left.\hspace{34mm}
+\sum_{k=1}^{N_f}\frac{1}{p^2+M^2_{kk}}\right).
\eeq
Here the $\alpha$ dependence has been reinserted. Note that the
first term in $\frac{1}{V}\sum_p G_{vv}$ is simply $\Delta_{vv}$ which
in the case of $\beta=2$ exactly cancels the propagator $\Delta_{vv}$
in (\ref{ccgenbeta}). Cancellations like this is a typical property
of $\beta=2$ and we shall see several additional examples of terms
proportional to the factor $\delta_\beta\equiv 1-2/\beta$, terms which
only
cancel in the case of $\beta=2$. Explicitly
\beq
\delta_\beta \equiv 1-\frac{2}{\beta} =
\left\{\begin{array}{rl}-1, &\beta=1\\0, &\beta=2\\ \frac{1}{2},
&\beta=4\end{array}\right.
\eeq
Taking advantage of the above we find that we can write the chiral
condensate
for general $\beta$ as
\beq
\frac{\Sigma^\beta(m_v,\{m\})}{\Sigma_0}
& = & 1-\frac{1}{F^2}\left(\vphantom{\sum_p}\frac{2}{\beta}N_f\Delta_{vs}
+\delta_\beta\Delta_{vv} \right. \nn \\
&& \left.\hspace{-17mm}-
\frac{1}{V}\sum_p\frac{(1+\delta_{\beta,1})(m_0^2+\alpha p^2)/N_c}
{(p^2+M_{vv}^2)(p^2+M_{vv}^2)
{\cal F}^{\beta}(p^2)}
 \right).
\label{ccgenbeta2}
\eeq


\section{Schwinger-Dyson equations and Virasoro constraints \label{vcsec}}
Having established the equivalence of the replica and the
supersymmetric methods in all classes of chiral symmetry breaking we
now turn towards calculating Schwinger-Dyson equations that govern the
behavior of the partition functions in the three classes of chiral
symmetry breaking. That is, we turn from (partially) quenched chiral
perturbation theory to ordinary chiral perturbation theory, but the
results are also useful for the quenched approximation. It turns out
that the Schwinger-Dyson equations can be expressed in a universal
way.

The method is here shortly outlined. For greater details see Appendix
\ref{sdeqapp}. When working with ordinary QCD ($\beta=2$) the finite
volume partition function is an integral over the unitary
group \cite{GL87,LS92}
\beq
\z{2}_\nu & = & \int_{U\in U(N_f)} \hspace{-7mm}dU (\det U)^\nu
e^{\frac{1}{2}
\tr(U{\cal M}^\dagger+U^\dagger{\cal M})}.
\label{part2}
\eeq
Here we have included the volume and the chiral condensate in ${\cal
M}=MV\Sigma_0$, where $M$ is the usual quark mass matrix.


The partition function is a periodic function in the vacuum angle,
$\theta$. This means that the full partition function ${\cal
Z}(\theta)$ may be expressed as a Fourier series, where the expansion
coefficients, ${\cal Z}_\nu$, correspond to the partition function in
sectors of fixed topological charge, $\nu$.

As discussed by Leutwyler and Smilga \cite{LS92}, the topological
charge need not be integer. Letting $n_L$ and $n_R$ denote the number
of left- and right-handed zero-modes of the covariant derivative
$\rlap {\hspace{-0.5mm} \slash} D$, respectively, the index theorem
states that
\beq
n_L-n_R & = & \frac{1}{16\pi^2}\int
d^4x\tr(G_{\mu\nu}\tilde G_{\mu\nu}).
\eeq
The trace depends upon the fermionic representation, and the trace of
the generators, $t^a$, is $\frac12\ell(r)\delta^{ab}=\tr[t^at^b]$.
$\ell(r)$ is the index of the fermionic representation.

The topological charge is
\beq
\nu & = & \frac{1}{32\pi^2}\int d^4x(G^a_{\mu\nu}\tilde G^a_{\mu\nu}),
\eeq
and thus it is possible to deduce the relation
\beq
\nu & = & \bar\nu\frac{1+\delta_{\beta,4}}{\ell(r)},
\label{nubarnu}
\eeq
where $\bar\nu$ is integer. The reason for the appearance of
$\delta_{\beta,4}$ in eq. (\ref{nubarnu}) is that for real
representations, the reality of the Dirac operator leads to all the
eigenfunctions of $\gamma_5$ appearing in pairs. In particular, the
number of zero-modes is even for $\beta=4$.

For the gauge groups and representations usually considered in the
literature the relation between $\nu$ and $\bar\nu$ is simple. For
$\beta=1$ one usually considers fermions in the fundamental
representation of $SU(N_c=2)$ for which $\ell(r)=1$ and
$\nu=\bar\nu$. The same is the case for ordinary QCD, the fundamental
representation of $SU(N_c=3)$. However, for $\beta=4$ the usual
scenario is adjoint (Majorana) fermions for which $\ell(r)=2N_c$ and
$\bar\nu=N_c\nu$ (again with the gauge group $SU(N_c)$). For
simplicity, these are also the only representations considered here,
but it is straightforward to generalize to other representations.


In the following, the crucial property of the partition function,
eq. (\ref{part2}), is that $(\det {\cal M})^{-\nu}\z{2}_\nu$ only
depends on the $N_f$ eigenvalues $x_i$, $i=1,\dots,N_f$, of ${\cal
MM}^\dagger$ \cite{DV01}.  This can be seen from rescaling ${\cal
M}\to V^{-1} {\cal M}$ and using the invariance of the unitary measure
under left and right multiplication.

This dependence on the eigenvalues of ${\cal MM}^\dagger$ only also holds
in the other classes of chiral symmetry breaking \cite{DV01}.
In the case of $\beta=1$ the symmetry breaking pattern is $SU(2N_f)\to
Sp(2N_f)$ and a convenient parametrization of the Goldstone fields is
$UIU^t$ with the partition function \cite{DV01}
\beq
\z{1}_\nu = \int_{U\in U(2N_f)}\hspace{-7mm} dU (\det U)^\nu
e^{\frac{1}{4}\tr((UIU^t)^\dagger
\tilde{\cal M}+UIU^t\tilde{\cal M}^\dagger)}.
\eeq
$I$ is the anti-symmetric unit matrix and $\tilde{\cal M}$ is an arbitrary
anti-symmetric complex matrix. Concentrating on theories without di-quark
terms, we have
\beq
I & = & \left[
\begin{array}{cc}
  0 & \mb{1} \\
  -\mb{1} & 0
\end{array}
\right], \hspace{3mm}
\tilde{\cal M} = \left[
\begin{array}{cc}
  0 & {\cal M} \\
  -{\cal M}^t & 0
\end{array}
\right],
\eeq
where we again define ${\cal M}=MV\Sigma_0$ with $M$ being the usual
$N_f\times N_f$ quark mass matrix. In this case $(\det\M)^{-\nu}\z{1}_\nu$
is a
symmetric function of the eigenvalues of $\tilde\M\tilde\M^\dag$
\cite{DV01}.

In the case of $\beta=4$ the Goldstone fields can be parametrized by
$UU^t$, $U\in U(N_f)$. With $\bar{\nu}\equiv N_c\nu$ the finite
volume partition function is \cite{LS92, SV95}
\beq
\z{4}_\nu = \int_{U\in U(N_f)} \hspace{-7mm} dU (\det U)^{2\bar{\nu}}
e^{\frac{1}{2}\tr((UU^t)^\dagger{\cal M}+UU^t{\cal M}^\dagger)},
\eeq
again with ${\cal M}=MV\Sigma_0$ but this time with $\cal M$ being a
symmetric matrix. $(\det\M)^{-\bar\nu}\z{4}_\nu$ is a
symmetric function of $\M\M^\dag$ \cite{DV01}.

Useful Schwinger-Dyson equations are obtained in Appendix
\ref{sdeqapp}.  We multiply the integrands in the partition functions
by $\tr t^aU\M^\dagger$ ($\beta=2$), $\tr t^aUIU^t\M^\dagger$
($\beta=1$) and $\tr t^aUU^t \M^\dagger$ ($\beta=4$), and then
subsequently act with the differential operator $\nabla^a$. The
generators of the integration manifold $t^a$ are normalized according
to $t^a_{ij}t^a_{kl}=\frac{1}{2}\delta_{il}\delta_{jk}$.  Since the
integral of a total derivative vanishes this gives relations among the
expectation values resulting in the Schwinger-Dyson equations
(\ref{sdeqapp2}), (\ref{sdeqapp1}), and (\ref{sdeqapp4}) given in the
appendix.

These equations can be written in a remarkably uniform way.  Note that
the number of generators, $\cal N$, of the cosets is 1 higher than the
dimension of the ``original'' cosets stated in the introduction
because of the projection onto sectors of fixed topological charge.
For instance for $\beta=2$ the integration goes from an integration
over $SU(N_f)$ to an integration over $U(N_f)$ by including the
topological charge. The dimensions of the ``original'' cosets are
stated in Table \ref{tablegen}.  Gathering the differential equations
deduced in Appendix \ref{sdeqapp} we see that it is possible to write
the differential equations universally as
\begin{widetext}
\beq \left[x_i^2\del_i^2+\left(\frac{{\cal N}}{N_f}+\nu\right)
x_i\del_i+\frac{2}{\beta}\sum_{\stackrel{i,j}{i\neq j}}
\frac{x_ix_j}{x_i-x_j}\left(\del_i-\del_j\right)\right]
\left(\prod_kx_k^{-\nu/2}\right){\cal Z}^\beta_\nu =
\frac{1}{4}\sum_ix_i\left(\prod_kx_k^{-\nu/2}\right){\cal
Z}^\beta_\nu,
\label{masterde}
\eeq
\end{widetext}
with $\nu\to\bar{\nu}$ for $\beta=4$.
This differential equation is one of the main results of the present
paper,
nicely combining the seemingly very different integrations over cosets
into
one governing differential equation in terms of the fermion masses. All
coset dependence is thus included in the single parameter $\beta$.
In writing this differential equation we have assumed that the mass matrix
$\cal M$ is real with positive eigenvalues allowing us to write
$(\det \M)^{-\nu}=\prod_ix_i^{-\nu/2}$.

In the following it will be convenient to introduce the notation
\beq
\frac{{\cal N}}{N_f} & = & \frac{2}{\beta}N_f+\delta_\beta.
\eeq

An alternative way of obtaining useful Schwinger-Dyson equations,
reminiscent of the one being presented here, is to insert a
$\delta_{ik}$ in the partition function in the form of
$U_{ij}U^\dag_{jk}$, $U\in U(N_f)$, subsequently realizing that this
can be reexpressed as a differential equation, the differentiation
being with respect to the mass matrices. However, one should realize
that when recognizing $\left<UU^\dag\right>$ as a differential
operator, in the cases of $\beta=1,4$ the fact that the mass matrices
are anti-symmetric and symmetric, respectively, causes complications
with regard to the diagonal elements. This is especially so for the
case of $\beta=4$, see for instance eq. (\ref{sdb4matrix2}). This
method of obtaining Schwinger-Dyson equations has for example been
employed in ref. \cite{DV01} to obtain the Virasoro constraints in the
small mass phase.  Thus, in eqs. (19) and (28) of ref. \cite{DV01} a
factor of 2 is implied for each differentiation with respect to a
diagonal element. Having deduced the differential equation
(\ref{masterde}), in which all dependence upon the mass matrices is
through the eigenvalues, we need no longer be concerned with this
point.

\begin{table}[ht]
$\renewcommand{\arraystretch}{2.5}
\begin{array}{|c|c|c|}
\hline
\beta & \mbox{Coset} & {\cal N}-1 \\
\hline
1 & SU(2N_f)/Sp(2N_f) & N_f(2N_f-1)-1 \\
2 & SU_L(N_f)\times SU_R(N_f)/SU(N_f) & N_f^2-1 \\
4 & SU(N_f)/SO(N_f) & N_f\frac{N_f+1}{2}-1 \\
\hline
\end{array}$
\caption{The number of generators of the cosets in the three classes of
chiral symmetry breaking.}
\label{tablegen}
\end{table}


\subsection{The small mass phase}
We now proceed to find Virasoro constraints for the finite volume
partition function by expanding in the eigenvalues of $\M\M^\dag$.
For the small mass phase this was first done in ref. \cite{MMS96} in
the simplest version, namely $\beta=2$ with vanishing topological
charge.  More recently, Virasoro constraints in the small mass phase
have been calculated in ref. \cite{DV01}. We present our calculation
of the small mass Virasoro constraints here mainly for two reasons. We
wish to demonstrate the ease by which we arrive at the Virasoro
constraints by using eq. (\ref{masterde}) and to demonstrate the
correctness of (\ref{masterde}) by exactly reproducing the results of
\cite{DV01}.

In the small mass phase a suitable set of expansion parameters for the
partition function is defined by
\beq
t_k & \equiv & \frac{1}{4^kk}\sum_ix_i^k,
\label{expsmallmass}
\eeq
which transforms the derivatives in (\ref{masterde}) into
\beq
\dif{x_i} & = & \sum_k\frac{1}{4^k}x_i^{k-1}\dif{t_k},\\
\difto{x_i^2} & = & \sum_k\frac{1}{4^k}(k-1)x_i^{k-2}\dif{t_k} \nn \\
& & + \sum_{k,l}\frac{1}{4^{k+l}}x_i^{k+l-2}\dif{t_k}\dif{t_l}.
\eeq
Note that using the invariance of the Haar measure as well as
permutation symmetry of the $x_i$, we are allowed to remove the sum
over $i$ \cite{BRT81}. Inserting the derivatives in our master
differential equation (\ref{masterde}) and removing the sum over $i$,
we obtain
\beq
&\left[\sum_k\frac{1}{4^k}(k-1)x_i^k\dif{t_k}
+\sum_{k,l}\frac{1}{4^{k+l}}x_i^{k+l}\dif{t_k}\dif{t_l} \right. &\nn \\
&\hspace{-8mm}+\left(\frac{2}{\beta}N_f+\delta_\beta+\nu\right)
\sum_k\frac{1}{4^k}x_i^k\dif{t_k}& \nn \\
&\hspace{-4mm}
+\frac{2}{\beta}\sum_{k=2}^{\infty}\sum_{l=1}^{k-1}(k-l)\frac{1}{4^l}x_i^l
t_{k-l}\dif{t_k}& \nn \\
&\left.\hspace{-7mm}
-\frac{2}{\beta}\sum_{k=2}^{\infty}(k-1)\frac{1}{4^k}x_i^k\dif{t_k}
-\frac{1}{4}x_i\right] \nn \\
& \hspace{-20mm}
\times \left(\prod_{i=1}^{N_f}x_i\right)^{-\nu/2}{\cal Z}^\beta_\nu
=0. &
\eeq
Determining the Virasoro constraints is now a matter of reading off
the coefficients of $x_i^k$. This procedure is justified provided the
coefficients are determined uniquely and in a consistent
way. Uniqueness is satisfied by requiring the boundary condition
${\cal Z}^\beta_\nu=1$ if all fermion masses vanish. Consistency of
the equations obtained by simply reading off the coeffecients of
$x_i^k$ is a more subtle point, all of the expansion coefficients
$t_k$ being independent only in the $N_f\to\infty$ limit. That this
procedure is consistent has been demonstrated in ref. \cite{MMS96}.
We thus find the constraints
\beq
\left[L_n^\beta-\frac{\beta}{2}\delta_{1,n}\right]
\left(\prod_i x_i\right)^{-\nu/2}\!\!{\cal Z}^\beta_\nu=0,
\eeq
with the Virasoro operators $L_n^\beta$
\beq
L^\beta_n & = &
\left(N_f+\frac{\beta}{2}(n\delta_\beta +\nu)\right)\dif{t_n}
+\frac{\beta}{2}\sum_{m=1}^{n-1}\dif{t_m}\dif{t_{n-m}} \nn \\
&&+\sum_{m=1}^{\infty}mt_m\dif{t_{m+n}}.
\label{virsmallmass}
\eeq
Again we have $\nu\to\bar\nu$ for $\beta=4$.  The normalization is
chosen such that the operators $L_n^\beta$ satisfy the Virasoro
algebra without central charge
\beq
[L_m^\beta,L_n^\beta] & = & (m-n)L^\beta_{m+n}, \hspace{5mm}m,n\geq 1.
\eeq
The virtue of the Virasoro operator formalism is that because the
operators satisfy the Virasoro algebra, in principle all coefficients
of a Taylor expansion in terms of the parameters (\ref{expsmallmass})
can be found from $L_1$ and $L_2$.  The Virasoro operators
(\ref{virsmallmass}) exactly match those found in ref. \cite{DV01}. For
the
solution to the constraints we also refer the reader to
ref. \cite{DV01}.


\subsection{The saddle-point approach to the large mass phase}
Virasoro constraints in the large mass phase for $\beta=2$ with
vanishing topological charge were found in ref. \cite{GN92} and solved
in ref. \cite{DS00A}. Below we will present the extension of these results
to include a non-vanishing topological charge.  We will also give the
first terms in a mass expansion of the $\beta=1,4$ partition functions
in the case of equal quark masses. Analytical results with equal quark
masses have been found in both of these symmetry breaking classes
\cite{SV95}. The disadvantage of these results is that they depend on
determinants proportional in size to $N_f$ which quickly makes
calculations somewhat cumbersome. The results presented here, on the
other hand, have a very simple $N_f$ dependence.

The difficulty in the large-mass expansion arises from the fact that
it is not possible to express the partition function in terms of a
simple Taylor expansion in the inverse masses.  However, as noted in
the literature \cite{GN92,MMS96}, for $\beta=2$ this becomes possible
after extracting a prefactor from the partition function. In this case
the prefactor turns out to be given by the saddle-point approximation,
or the classical contribution, of the partition function and thus one
would expect this to be the case in general. The classical
contribution is defined as the $N_f\to\infty$ limit. We have been able
to determine the classical contribution to the next to leading order
only in the case of $\beta=2$. For $\beta=2$ this is the order needed
to be able to include the topological charge in the Virasoro
constraints.  In the case of $\beta=1,4$ however, we have found
indications that the next to leading order term in the saddle-point
approximation is itself given by an infinite mass-expansion. Even so,
using the lowest order correction in this mass-expansion we have
calculated the lowest order mass dependence of the partition function
for $\beta=1,4$ with equal fermion masses and non-vanishing
topological charge. Using this result we calculate the one-loop
correction to the quenched chiral condensate.

Following \cite{BN81,GN92} we define the free energy, $F$, in sectors of
vanishing topological charge by ${\cal Z}^\beta_0
=e^{N_fF}$. Defining in general
${\cal Z}^{\beta}_\nu=e^{N_fF_\nu}$ we find it
useful to introduce
\beq
G_\nu & = & F_\nu -\frac{1}{N_f}\frac{\nu}{2}\sum_i \ln x_i,
\eeq
since then $(\prod_i x_i^{-\nu/2}){\cal Z}^{\beta}_\nu=e^{N_f G_\nu}$. For
the
differential operators this means
\beq
\dif{x_i}e^{N_f G_\nu} & = & N_fe^{N_f G_\nu}\dif{x_i}G_\nu \\
\difto{x_i^2} e^{N_f G_\nu} & = & N_fe^{N_f G_\nu}
\left[N_f\left(\frac{\del G_\nu}{\del x_i}\right)^2 \!\!\!
+\frac{\del^2 G_\nu}{\del x_i^2}\right].
\eeq
In what follows we will use the somewhat sloppy terminology that $G_\nu$
is the free energy.

The free energy, being an evaluation around the $N_f\to\infty$
saddle-point, can be written as a power expansion in $1/N_f$.  When
doing the counting of orders of $N_f$ the action counts as order $N_f$.
Explicitly including this factor to ease the counting of powers,
when inserting the free energy into the differential equation
(\ref{masterde}) we find the equation
\beq
0 & = & \sum_i\left[\vphantom{\sum_{\stackrel{j}{j\neq
i}}}x_i^2\left\{\left(\frac{\del
G_\nu}{\del x_i}\right)^2+ \frac{1}{N_f}\difto{x_i^2}G_\nu\right\}
\right. \nn \\ &&\left.
+x_i\left\{\frac{2}{\beta}+\frac{\delta_\beta+\nu}{N_f}\right\}\dif{x_i}G_\nu
\nn \right. \\ && \left.  +\frac{2}{\beta N_f}\sum_{\stackrel{j}{j\neq
i}}
\frac{x_ix_j}{x_i-x_j}\!\left(\frac{\del G_\nu}{\del x_i}
-\frac{\del G_\nu}{\del x_j}\right)\!-\!\frac{1}{4}x_i\right].
\label{freede}
\eeq
In refs. \cite{BN81,GN92,MMS96}
the classical solution for $\beta=2$ with $\nu=0$ was found to be
\beq
\sum_i\sqrt{x_i}
-\frac{1}{2N_f}\sum_{i,j}\ln(\sqrt{x_i}+\sqrt{x_j}).
\eeq
Ignoring the second derivative and $\delta_\beta$, both of which are
suppressed by an order of $N_f$, we likewise find the solution to
(\ref{freede}) to lowest order in $1/N_f$
\beq
\sum_i\sqrt{x_i}
-\frac{1}{\beta N_f}\sum_{i,j}\ln(\sqrt{x_i}+\sqrt{x_j}).
\label{classicalfe}
\eeq
From studies of the partition function for small $N_f$ and with equal
masses,
using the analytical expressions of ref. \cite{SV95}
we know that (\ref{classicalfe}) is
too simple; we expect that $\delta_\beta$ has to be included in the
prefactor. We take care of this by
expanding the free energy as
\beq
G_\nu & = & G_\nu^{(0)}+\frac{1}{N_f}G_\nu^{(1)}+\dots
\eeq
where $G_\nu^{(0)}$ is defined by (\ref{classicalfe}).  Inserting this
expansion in (\ref{freede}) we now find the following differential
equation for $G_\nu^{(1)}$
\beq
0 & = & \frac{1}{N_f}\sum_i\left[\vphantom{\sum_{\stackrel{j}{j\neq i}}}
\sqrt{x_i}\frac{\delta_\beta+2\nu}{4}
-\frac{\delta_\beta}{2\beta N_f}
\sum_j \frac{\sqrt{x_ix_j}}{(\sqrt{x_i}+\sqrt{x_j})^2}
\nn \right.\\&&\left.-
\frac{\nu}{\beta N_f}
\sum_j \frac{\sqrt{x_i}}{\sqrt{x_i}+\sqrt{x_j}}
\right. \nn \\
&& + \left. x_i^{3/2}\dif{x_i}G_\nu^{(1)} + \frac{2}{\beta N_f}\sqrt{x_i}
\sum_j \frac{\sqrt{x_ix_j}}{\sqrt{x_i}+\sqrt{x_j}}\dif{x_i}G_\nu^{(1)}
\right. \nn \\
&& \left. +\frac{2}{\beta N_f}
\sum_{\stackrel{j}{j\neq i}}\frac{x_i x_j}{x_i-x_j}\left(\dif{x_i}
-\dif{x_j}\right)G_\nu^{(1)}
\right],
\label{g1de}
\eeq
where we note that $G_\nu^{(1)}$ vanishes for $\beta=2$ with vanishing
topological charge since $\delta_2=0$.

At this point a few comments are in order. All terms in (\ref{g1de})
have to cancel order by order in $x$, $x\in x_i, i=1,\dots,N_f$. Thus
$\partial_iG_\nu^{(1)}$ only has terms of order less than or equal to
${\cal O}(x^{-1})$. Otherwise cancellations would only happen if
$\partial_iG_\nu^{(1)}$ had terms of all positive orders of $x$. From
this consideration it is clear that the term of order $\sqrt{x}$ in
(\ref{g1de}) should be cancelled by the term of order
$x^{3/2}\partial_iG_\nu^{(1)}$. This provides a condition for
$G_\nu^{(1)}$ which should then be inserted in (\ref{g1de}). If this
expression fails to cancel the remaining terms we conclude that
$\partial_iG_\nu^{(1)}$ also has an $x^{-3/2}$ part and the process
starts over again. Hopefully the process terminates after a finite
number of iterations.

From these considerations we conclude that
\beq
\dif{x_i}G_\nu^{(1)} & = & -\frac{\delta_\beta+2\nu}{4x_i}+\dots \nn \\
\Rightarrow \hspace{1cm} G_\nu^{(1)} & = & -\frac{\delta_\beta+2\nu}{4}
\sum_j\ln{x_j}+\dots,
\label{firstiteration}
\eeq
where in the last step we used the fact that $G_\nu^{(1)}$ should be a
symmetric function of the $x_i$.  Unfortunately the iteration process
does not stop within the first three levels of iteration. Indeed, the
results become increasingly complicated - thus we choose to stop the
iteration at this point, using (\ref{firstiteration}) as our only
correction to (\ref{classicalfe}). It turns out that this is enough to
calculate the partition function if one requires the fermions to have
equal masses. However, since all remaining terms are proportional to
$\delta_\beta$, (\ref{firstiteration}) is precisely sufficient to
calculate Virasoro constraints in the $\beta=2$ case with
non-vanishing topological charge. Furthermore, the remaining terms are
independent of $\nu$. Thus, in all classes of symmetry breaking, to
this order the $\nu$ dependence of the free energy is simply given by eq.
(\ref{firstiteration}).

First we consider $\beta=2$ with non-vanishing topological charge.
We write the partition function as
\beq
(\prod_i x_i^{-\nu/2}) \z{2}_\nu & = &
e^{N_f(G_\nu^{(0)}+\frac{1}{N_f}G_\nu^{(1)})}Y(\{t_k\}) \nn \\
& = &
e^{\sum_i\sqrt{x_i}}\prod_{i,j}\frac{1}{\sqrt{\sqrt{x_i}+\sqrt{x_j}}}
\nn \\
&& \times\prod_l{x_l^{-\nu/2}}Y(\{t_k\}).
\label{partbeta2top}
\eeq
In the second line we have removed the explicit $N_f$-dependence of
the action, and returned to our original normalization. We define the
set of variables relevant to a large mass expansion differently from
the small-mass expansion
\beq
t_k & = & -\frac{2^{2k+1}}{2k+1}\tr{\left( (\M\M^\dag)^{-k-1/2}\right)}
\nn \\
& = & -\frac{2^{2k+1}}{2k+1}\sum_i x_i^{-k-1/2},
\eeq
with the normalization chosen for convenience. Inserting
(\ref{partbeta2top}) in the governing differential equation, eq.
(\ref{masterde}) and reading off the coefficients of powers of $x_i$
as in the small mass expansion we find the Virasoro constraints
($\nu\to\bar\nu$ for $\beta=4$)
\beq
L_{n,\nu}^{\beta=2}Y(\{t_k\}) & = & 0, \hspace{5mm} n > 0,
\eeq
with the Virasoro operators
\beq
L_{0,\nu}^{\beta=2} & = & \sum_{k=0}^\infty (k+1/2)t_k\dif{t_k}+
\frac{1-4\nu^2}{16}+\dif{t_0} \\
L_{n,\nu}^{\beta=2} & = & \sum_{k=0}^\infty (k+1/2)t_k\dif{t_{k+n}}+
\frac{1}{4}\sum_{k=1}^n \difto{t_{k-1}\del t_{n-k}} \nn \\
&& +\dif{t_n},
 \hspace{5mm} n\geq 1.
\eeq
Again these operators satisfy the Virasoro algebra without central
charge.  Uniqueness of the solution is ensured by the boundary
condition $Y(\{t_k\})=1$ for all expansion coefficients
vanishing (the limit of infinite fermion masses).  Comparing these
Virasoro
operators with the Virasoro operators obtained in ref. \cite{GN92} we
see that the inclusion of a non-vanishing topological charge only
changes the Virasoro operator $L_0^{\beta=2}$. However, this still has
profound consequences for the power expansion of the partition
function, as can be readily checked.

To determine the first corrections due to including a non-vanishing
topological charge, we proceed by expanding the function $Y$ as
\beq
Y(\{t_k\}) = 1+\sum_{k=0}^\infty c_kt_k+
\!\sum_{0\leq k_1 \leq k_2} \! c_{k_1,k_2}t_{k_1}t_{k_2}+\dots.
\label{yexpansion}
\eeq
The coefficients are easily obtained from the Virasoro constraints and we
find to fourth order in the inverse masses
\beq
&& \!\!\!\z{2}_\nu = e^{\sum_i\sqrt{x_i}}\prod_{i,j}
\frac{1}{\sqrt{\sqrt{x_i}+\sqrt{x_j}}} \nn \\
& & \times\left[1+\frac{1-4\nu^2}{8}\tr(\M\M^\dag)^{-1/2}\right. \nn \\
& & +\frac{(1-4\nu^2)(9-4\nu^2)}{128}(\tr(\M\M^\dag)^{-1/2})^2 \nn \\
& &
+\frac{(1-4\nu^2)(9-4\nu^2)(17-4\nu^2)}{3\cdot1024}(\tr(\M\M^\dag)^{-1/2})^3
\nn \\
& & +\frac{(1-4\nu^2)(9-4\nu^2)}{3\cdot128}\tr(\M\M^\dag)^{-3/2} \nn \\
& & +\frac{(1-4\nu^2)(9-4\nu^2)(17-4\nu^2)(25-4\nu^2)}{3\cdot32768} \nn \\
& & \:\:\:\times(\tr(\M\M^\dag)^{-1/2})^4 \nn \\
& & +\frac{(1-4\nu^2)(9-4\nu^2)(25-4\nu^2)}{3\cdot1024} \nn \\
& & \left.\:\:\:\times\tr(\M\M^\dag)^{-1/2}
\tr(\M\M^\dag)^{-3/2} + \dots\right].
\label{partbeta2nu}
\eeq
The partition function for $\beta=2$ is known in closed form to be
\cite{JSV97}
\beq
{\cal Z}^{\beta=2}_\nu & = & \frac{\det A(\{\sqrt{x}\})}{\Delta(\{x\})}.
\label{closedpart}
\eeq
$A$ is an $N_f\times N_f$ matrix defined by
\beq
A(\{\sqrt{x}\})_{ij} & = & x_i^{(j-1)/2}I_{\nu+j-1}(\sqrt{x_i}),
\eeq
with $I_n$ a modified Bessel function. $\Delta$ is the Vandermonde
determinant
\beq
\Delta(\{x\}) & \equiv & \prod_{i>j}^{N_f}(x_i-x_j) \nn \\
& = & \det[(x_i)^{j-1}].
\eeq
Expanding (\ref{closedpart}) for $N_f=2$, which is sufficient to reveal
most subtleties, we find complete agreement between
this closed expression and our expansion, equation (\ref{partbeta2nu}).

Turning now to the other classes of chiral symmetry breaking, we find the
partition function for equal fermion masses. This is performed as before,
by writing
${\cal Z}^\beta_\nu$ as the prefactor times a power expansion in the
masses,
that is
\beq
{\cal Z}^\beta_\nu\!=\!
e^{\sum_i\sqrt x_i}\!
\prod_{i,j}\frac{1}{(\sqrt x_i \! + \! \sqrt x_j)^{1/\beta}}
\prod_kx_k^{-\delta_\beta/4}Y(\{t_k\}).
\eeq
Inserting this
expansion in
(\ref{masterde}) and setting all fermion masses equal, after some algebra,
we
obtain
\beq
-2c_0 & = & \frac{1+(N_f-1)2\delta_\beta/\beta-4\nu^2}{8},
\label{firstexpansion}
\eeq
where, as expected, we find that the partition functions are equal for
$N_f=1$, a feature which should be present not only to this order in
the mass expansion but to all orders. We have here returned to our
original normalization, removing the explicit $N_f$ dependence from
the partition function.  Seeing how the $\nu$ dependence of the free
energy was found to be very simple to the order examined, it should
not come as a complete surprise that the $\nu$-dependence is
independent of $\delta_\beta$, apart from the difference in
prefactors.  Explicitly writing the partition function for equal
masses to lowest order (defining $x\equiv x_1=\dots=x_{N_f}$)
\beq
{\cal Z}_\nu^\beta \!& = & \!e^{N_f\sqrt x}x^{-N_f^2/2\beta}x^{-N_f\delta
_\beta/4}\nn \\
&& \!\times\!\left(\!1\!+\!\frac{1+(N_f-1)2\delta_\beta/\beta-4\nu^2}{8}
\frac{N_f}{\sqrt x}\!+\!\dots\!\right)\!\!,
\eeq
where we as usual let $\nu\to\bar\nu$ for $\beta=4$.

This result is just what we need to calculate the finite volume
quenched chiral condensate in the large mass limit. To this end we
replace $N_f$ with $N_v$ or, formally, extend the partition function
with one replica set of $N_v$ fermions and then let $N_f\to 0$.
Changing variables to $\mu_v\equiv \sqrt x$ (with the
equal fermion masses given by $m_v=\mu_v/\Sigma_0V$)
eq. (\ref{cc}) can also be written
\beq
\frac{\Sigma_\nu^\beta(\mu_v)}{\Sigma_0} & = & \lim_{N_v\to 0}
\frac{1}{N_v}\dif{\mu_v}\ln{\cal Z}_\nu^\beta.
\eeq
We thus calculate the quenched chiral condensate in sectors of topological
charge $\nu$ ($\nu\to\bar\nu$ for $\beta=4$) and in all classes of
chiral symmetry breaking
\beq
\left. \frac{\Sigma_\nu^\beta(\mu_v)}{\Sigma_0}\right|_{\mbox{\scriptsize
quenched}}
\!\!\!\!\!= 1-\frac{\delta_\beta}{2\mu_v}
-\frac{1-2\delta_\beta/\beta-4\nu^2}{8\mu_v^2}+\dots.
\label{cclargemass}
\eeq
While the quenched chiral condensate has been previously calculated
in the case of $\beta=2$ \cite{D01}, interestingly for $\beta=1,4$ our
results show that we have to include a term of order ${\cal
O}(\mu_v^{-1})$ in these classes. This feature turns out to be very
important in the matching of the chiral condensate in the two finite
volume perturbation schemes considered in the next section.


\section{Quenched finite volume chiral condensates \label{sec:evsp}}
So far everything we have calculated has been in the large, albeit
finite, volume, in which the usual momentum expansion is
applicable. Now we turn to the other finite volume regime, in which
the correlation lengths of the Goldstone modes are much larger than
the volume of the box. In this regime the usual $p$-expansion of
chiral perturbation theory breaks down due to the propagation of
zero-momentum Goldstone bosons. Thus another expansion scheme, known
as the $\epsilon$-expansion \cite{GL87, HL90}, is required. In
ref. \cite{D01} it was shown, in the case of $\beta=2$, that a region
exists in which the two expansion schemes overlap. The main result of
this section is that matching is not fortuitous, the result carries
over to the $\beta=1,4$ cases. We demonstrate this by calculating the
quenched chiral condensate in the $\epsilon$-expansion and comparing
with the corresponding result (\ref{ccgenbeta2}) of the
$p$-expansion. Thus we conclude that matching of the two expansion
schemes is possible in all classes of chiral symmetry breaking.

Again we use the ${\cal O}(p^2)$ Lagrangian which we parameterize as in
(\ref{lagrangian}), i.e.
\beq
{\cal L}_{\mbox{eff}} & = &
\frac{F_\beta^2}{4}\tr(\partial_\mu U^\dag\partial_\mu U) -
\frac{\Sigma_0}{2}\tr({\cal M}^{(\beta)}(U+U^\dag)) \nn \\
& & + \frac{m_0^2}{2N_c}\Phi_0^2+
\frac{\alpha}{2N_c}\partial_\mu\Phi_0\partial_\mu\Phi_0.
\eeq

The virtue of the $\epsilon$-expansion is that it counts zero-momentum
modes as order ${\cal O}(1)$ while non-zero modes count as higher
orders in $\epsilon$.  Letting the four-volume be $V\equiv L^3/T$ and
treating the fermion masses as small in comparison with $T$ and $1/L$
one fixes the counting of orders of $\epsilon$ by setting $1/L={\cal
O}(\epsilon)$. The fermion masses $m_i$ are then counted as ${\cal
O}(\epsilon^4)$ which implies $M_{ij}\sim{\cal O}(\epsilon^2)$. Thus
all graphs which exclusively involve zero momentum propagators are
counted as being of order ${\cal O}(\epsilon^0)$ \cite{GL87}, see for
instance eq. (\ref{delta}). In the unquenched theory, adding a mass
term for the non-zero modes provides a Gaussian damping factor for
these modes which ensures that the non-zero modes count as ${\cal
O}(\epsilon^1)$ \cite{GL87}. In the quenched theory the counting is
not as simple since the mass term $m_0$ is a free parameter. But
following \cite{D01,DDHJ02} we perform a simultaneous expansion in
$\epsilon$ and $m_0^2/(N_cF^2)$, as a way of avoiding having to
evaluate all orders of $m_0$ for each order of $\epsilon$.

The idea, adopted from ref. \cite{GL87}, is to collect the
zero-momentum modes in the constant matrix $U_0\in G/H$ and write
\beq
U(x) & = & U_0 e^{ i\sqrt{2}\xi(x)/F_\beta},
\label{cov}
\eeq
where the field $\xi(x)\in G/H$ contains the non-zero momentum modes.
For the non-zero modes we choose the same parameterization as in
(\ref{fieldbeta1}), for $\beta=1$, and (\ref{fieldbeta4}), for
$\beta=4$.  That is ($\beta=1$)
\beq
\xi = \frac{1}{\sqrt{2}}\left(\ba{cc} \phi & \psi \\
\psi^\dagger & \phi^T\ea\right),
\eeq
with $\phi$ Hermitian and $\psi$ anti-symmetric complex fields, and
($\beta=4$)
\beq
\xi=\left(\begin{array}{cccc}
A_{11} & \frac{1}{\sqrt{2}}A_{12} & \cdots &
\frac{1}{\sqrt{2}}A_{1,N_f+N_v} \\
\frac{1}{\sqrt{2}}A_{12} & \ddots& & \vdots\\
\vdots& & \ddots & \vdots \\
\frac{1}{\sqrt{2}}A_{1,N_f+N_v}& \cdots& \cdots & A_{N_f+N_v,N_f+N_v}
\end{array}\right),
\eeq
with $\xi$ being a real, symmetric matrix.

We now wish to perform this change of variables and integrate over the
zero modes and the non-zero momentum modes separately. Thus we have to
perform the change of variables (\ref{cov}) in the partition
function. The Jacobian of this change of variables is, to lowest order,
just a constant, corresponding to a shift of the vacuum
energy. Noting that integrals of odd
powers of $\xi$ vanish, with this change of variables the lowest order
action
becomes
\beq
S \!& = & \!\int d^4x \tr
\left[ \frac{1}{2}\del_\mu \xi (x)\del_\mu \xi(x) \right. \nn \\
&& \!\left. +\frac{m_0^2}{2N_c}
\left(\Xi_0+\Xi(x) \right)^2
+  \frac{\alpha}{2N_c}\del_\mu\Xi(x)\del_\mu\Xi(x)\right] \nn \\
&&\!
-\frac{\Sigma_0}{2}\tr\left[{\cal M}(U_0+U_0^\dag)\left(V\!-\!
\frac{1}{F_\beta^2}\int\!\! dx \xi^2
\right)\right]\!\!,
\eeq
where $U_0\equiv e^{i\sqrt{2}\varphi_0/F_\beta}$, and
$\Xi_0\equiv\tr\varphi_0$.  The mass matrices $\cal M$ are defined as in
eqs. (\ref{massbeta1}) and (\ref{massbeta4}).

Since we are once again interested in calculating the one loop
correction to the chiral condensate we integrate out the non-zero
momentum modes, $\xi(x)$. To this end we define the non-zero momentum
modes
\beq
\bar{\Delta}_{ij} & \equiv &
\frac{1}{V}\sum_{p\neq0}\frac{1}{p^2+M_{ij}^2}\\
\bar{G}_{ij} & \equiv & \frac{1}{V}\sum_{p\neq0}G(M_{ij}^2).
\eeq
The two-point functions for the off-diagonal mesons are
related to the full propagators by
\beq
\bar{\Delta}_{ij} & = & \Delta_{ij}-\frac{1}{VM_{ij}^2}.
\eeq
The propagator of the diagonal mesons, $G(M_{ij}^2)$, is defined as in
eq. (\ref{diagprop}).

Thus integrating out the fluctuating fields we find the fully quenched
one loop correction to the effective Lagrangian for the zero momentum
modes
by putting $N_f\to 0$,
\beq
&&\lim_{N_v\to 0}\frac{1}{N_v}
\left<\frac{\Sigma_0}{2F_\beta^2}\tr\left[{\cal M}(U_0+U_0^\dag)\left(\!
\int dx \xi(x)^2
\right)\right]\right> \nn \\
&=& \lim_{N_v\to 0}\frac{1}{N_v}
\frac{V\Sigma_0}{2F^2}\tr({\cal M}(U_0+U_0^\dag)) \nn \\
&& \hspace{10mm}\times\left(
\frac{2}{\beta}
(N_v-1)\bar{\Delta}_{vv}+\bar{G}_{vv}\right) \nn \\
& = & \lim_{N_v\to 0}\frac{1}{N_v}\frac{V\Sigma_0}{2F^2}\tr({\cal
M}(U_0+U_0^\dag)) \nn \\
&& \hspace{10mm}
\times\left(\delta_\beta\bar{\Delta}_{vv}-
(1+\delta_{\beta,1})\frac{\alpha}{N_c}\bar{\Delta}_{vv}+\right. \nn
\\&& \hspace{14mm}\left.\frac{1}{N_c}
(1+\delta_{\beta,1})(m_0^2-\alpha
M^2_{vv})\del_{M^2_{vv}}\bar{\Delta}_{vv}
\right)\!.
\label{finitecorrection}
\eeq
In the last equality we have made use of the fact that the first part of
the
diagonal propagator is the same as for the off-diagonal propagator. That
is
\beq
\bar{G}_{ij} = \bar{\Delta}_{ij}-\sum_{p\neq0}\frac{(\delta_{\beta,1}+1)
(m_0^2+\alpha p^2)/N_c}
{(p^2+M_{ii}^2)(p^2+M_{jj}^2){\cal F}^\beta(p^2)}.
\eeq
Equation (\ref{finitecorrection}) is very useful, since from the
correction to the zero'th order approximation
we see that the one-loop improved effective partition function
can be written in the same way as the
zero'th order approximation
\beq
{\cal Z}^\beta & = & \int_{G/H}dU_0 e^{\frac{1}{2}V\Sigma_0\tr{\cal M}
(U_0+U_0^\dag)-\frac{Vm_0^2}{2N_c}\Xi_0^2},
\label{zeroth}
\eeq
by simply changing the mass eigenvalues into
\beq
\mu_v' & = & \mu_v\left[1-\frac{1}{F^2N_c}
\left\{N_c\delta_\beta\bar\Delta_{vv}
-(1+\delta_{\beta,1})\alpha\bar{\Delta}_{vv}
\right.\right.\nn
\\&& \left.\left.\hspace{5mm}+
(1+\delta_{\beta,1})(m_0^2-\alpha
M^2_{vv})\del_{M^2_{vv}}\bar{\Delta}_{vv}
\right\}\vphantom{\frac{1}{F^2N_c}}\right].
\label{muchange}
\eeq
Notice that we again define $\mu_v\equiv m_v V\Sigma_0$.

Thus we can exploit the usual formula, (\ref{cc}), to find the one
loop improved quenched chiral condensate
\beq
\frac{\Sigma(\mu_v)}{\Sigma_0} & = & \lim_{N_v\to 0}\frac{1}{N_v}
\frac{\del \mu'_v}{\del \mu_v}\dif{\mu'_v}\ln{{\cal Z}^\beta(\mu'_v)}.
\label{finqcc}
\eeq

Now we make use of our calculation of the chiral condensate in sectors
of fixed topological charge. When calculating the chiral condensate, we
have
to sum over topology \cite{D99}. To perform this summation we follow
the idea of ref. \cite{D01} and include an arbitrary vacuum angle in
the partition function by absorbing it into the field $\Xi_0$. Next we
perform a Fourier transform projecting the vacuum angle onto sectors
of topological charge.

Identification of the topological susceptibility \cite{OTV99} is performed
in Appendix \ref{topsusapp}. The result is
\beq
\left<\nu^2\right> & = & (1+\delta_{\beta,1})\frac{m_0^2VF^2}{2N_c}.
\eeq
The summation over topologies is
\beq
\Sigma^\beta(\mu_v)
= \left<\Sigma^\beta_\nu(\mu_v)\right>
= \sum_{\nu=-\infty}^{\infty}\frac{{\cal Z}^\beta_\nu(\mu_v)
\Sigma^\beta_\nu(\mu_v)}{{\cal Z}^\beta(\mu_v)}.
\eeq
Inserting the quenched chiral condensate in sectors of topological charge
calculated in eq. (\ref{cclargemass}), to the order considered we find
\beq
\frac{\Sigma^\beta(\mu_v)}{\Sigma_0} & = &
\left<\frac{\Sigma_\nu^\beta(\mu_v)}{\Sigma_0}\right> \nn \\
& = & 1-\frac{\delta_\beta}{2\mu_v}
-\frac{1-2\delta_\beta/\beta-4\left<\nu^2\right>}{8\mu_v^2}.
\label{avsigma}
\eeq

Now we look for a size of the volume $V$ in which the two finite volume
regimes match \cite{D01}.
Obviously (\ref{avsigma}) cannot be a simple perturbation of the
infinite volume theory, since $\mu_v\equiv m_vV\Sigma_0$
in this regime is finite
while $\left<\nu^2\right>\sim V$. Thus the requirement that the last term
in
(\ref{avsigma}) is a small perturbation is
\beq
V & \gg & (1+\delta_{\beta,1})\frac{F^2m_0^2}{4N_cm_v^2\Sigma_0^2}.
\eeq
The requirement of this extreme finite volume regime is that $V\ll
M_{vv}^{-4}= F^4/4m_v^2\Sigma_0^2$. Both limits can be met by
letting $N_c\to\infty$ while
keeping all other quantities fixed, since then $F^2\gg
(1+\delta_{\beta,1})m_0^2/N_c$.

Having established a region in which a match is possible, let us
first review the result (\ref{ccgenbeta2})
for the chiral condensate in the large, but finite,
volume. Rewriting this for the quenched limit we obtain
\beq
\frac{\Sigma^\beta(\mu_v)}{\Sigma_0} & = &
1-\frac{1}{F^2N_c}
\left\{N_c\delta_\beta\Delta_{vv}
-(1+\delta_{\beta,1})\alpha{\Delta}_{vv}
\right.\nn
\\&& \left.\hspace{2mm}+
(1+\delta_{\beta,1})(m_0^2-\alpha M^2_{vv})\del_{M^2_{vv}}{\Delta}_{vv}
\right\}.
\label{finres}
\eeq

In the extreme finite volume we also have all the ingredients
necessary for calculating the quenched chiral condensate. Inserting
eqs. (\ref{muchange}) and (\ref{avsigma}) in eq. (\ref{finqcc}) we
find to leading order
\beq
\frac{\Sigma^\beta(\mu_v)}{\Sigma_0} & = &
\left(1-\frac{\delta_\beta}{VF^2M^2_{vv}}+
(1+\delta_{\beta,1})\frac{m_0^2}{N_cF^2M_{vv}^4V}\right) \nn \\
&& \times \left[1-\frac{1}{F^2N_c}
\left\{N_c\delta_\beta\bar\Delta_{vv}
-(1+\delta_{\beta,1})\alpha\bar{\Delta}_{vv}
\right.\right.\nn
\\&& \left.\left.\hspace{5mm}+
(1+\delta_{\beta,1})(m_0^2-\alpha
M^2_{vv})\del_{M^2_{vv}}\bar{\Delta}_{vv}
\right\}\vphantom{\frac{1}{F^2N_c}}\right] \nn \\
& = &
1-\frac{1}{F^2N_c}
\left\{N_c\delta_\beta\Delta_{vv}
-(1+\delta_{\beta,1})\alpha{\Delta}_{vv}
\right.\nn
\\&& \left.\hspace{5mm}+
(1+\delta_{\beta,1})(m_0^2-\alpha M^2_{vv})\del_{M^2_{vv}}{\Delta}_{vv}
\right\}.
\label{vfinres}
\eeq
In this calculation, we have made use of the fact that
$\left<\nu^2\right>\gg 1$ while $\mu_v\sim {\cal O}(1)$.

Comparing the results (\ref{vfinres}) for the extreme finite volume
and (\ref{finres}) for the ordinary finite volume, we find the
promised overlap of the two perturbative schemes. This is an important
consistency check on our calculations and it is a very attractive feature
of this calculation that it is precisely the term proportional to
$1/\mu_v$ in the quenched chiral condensate, absent in the $\beta=2$
case, which makes this matching possible.


\section{Conclusions}
We have analyzed the three patterns of chiral symmetry breaking
relevant to chiral perturbation theory, in particular $\beta=1$ and
$\beta=4$ corresponding to fermions in a pseudo-real and real
representation, respectively.

In the usual $p$-expansion of chiral perturbation theory, we have
demonstrated how the replica method is equivalent to the
supersymmetric method by comparing the Feynman rules to one loop order
for the $\beta=1,4$ classes of chiral symmetry breaking, and seeing
how the calculations proceed it is fairly obvious how this generalizes
to higher order loop calculations.

Next, using Schwinger-Dyson equations, we have obtained a governing
Schwinger-Dyson equation for the partition function in all classes of
chiral symmetry breaking and in sectors of fixed topological
charge. We have demonstrated how it is possible to write this
differential equation in terms of the mass eigenvalues only, with all
dependence upon the class of chiral symmetry breaking being through
the Dyson index, $\beta$.

We have utilized the governing differential equation to determine
Virasoro constraints in the small mass expansion, which exactly
matches those derived in ref. \cite{DV01}. We have also calculated
Virasoro constraints in the large mass expansion of the $\beta=2$
partition function in sectors of fixed topological charge. We note
how the inclusion of the topological charge in the Virasoro
constraints becomes possible by simply expanding the saddle-point
contribution of the free energy to next to leading order in $1/N_f$
instead of only keeping leading order terms.

We find indications of how the next to leading order of the
saddle-point approximation for $\beta=1,4$ seems to consist of an
infinite series in the fermion masses. However, we note that it is
still possible to find the first terms in a power expansion of the
partition function in these classes of chiral symmetry breaking if one
puts the masses equal. Having thus obtained the equal-mass partition
function in sectors of fixed topological charge to the order needed,
we calculate the quenched chiral condensate. We find that the
mass-dependence of the quenched chiral condensate is different for
real and pseudo-real representations of the fermions compared with the
formerly known result for fermions in complex representations.

In the other relevant finite volume, the volume smaller than the
correlation lengths of the Goldstone bosons, we utilize the
$\epsilon$-expansion to calculate the first correction to the quenched
chiral condensate. It turns out that it is exactly the above mentioned
difference in mass-dependence of $\beta=1,4$ which makes the quenched
chiral condensate of this finite volume match the corresponding
quenched chiral condensate of the ordinary $p$-expansion. Thus we find
a region in which the two perturbative expansions of chiral
perturbation theory, the $\epsilon$-expansion and the $p$-expansion,
precisely match.


\begin{acknowledgments}
The author wishes to thank P.H. Damgaard and K. Splittorff for useful
discussions.
\end{acknowledgments}


\appendix
\section{Schwinger-Dyson equations \label{sdeqapp}}
In this appendix we will find Schwinger-Dyson equations in all three
classes of chiral symmetry breaking. Theories with complex fermions,
such as ordinary QCD, have been considered before \cite{BN81}, but
since the calculations here are basically a less complicated version
of the calculations in the $\beta=1,4$ cases, we will line them up.
Also, this will be done in a somewhat different way from the
literature.

To avoid an unnecessarily complicated notation we consider first
theories with vanishing topological charge. The topological charge is
easily included at the end of the calculations.

\subsection{$\beta=2$}
We start out with the partition function in sectors of vanishing
topological
charge
\beq
\z{2} & = & \int_{U\in U(N_f)}dU \: e^{\frac{1}{2}
\tr\left(\M U^\dagger+{\M}^\dagger U\right)}.
\label{partapp2}
\eeq
As discussed in section \ref{vcsec}
the crucial property of this partition function is that
it only depends on the $N_f$ eigenvalues $x_i$ of ${\M}
{\M}^\dag$. To arrive at the Schwinger-Dyson equations,
we utilize the fact that
the integral of a total derivative vanishes. Thus
\beq
0 & = & \int_{U\in U(N_f)}dU\tr t^a\nabla^a\left(F(U)e^{\frac{1}
{2}\left({\M}U^\dagger+{\M}^\dagger U\right)}\right) \nn \\
& = & \left<\tr t^a\nabla^aF(U) \vphantom{\frac{1}{2}}\right.\nn \\
& & \hspace{3mm}\left. +\frac{1}{2}\tr (t^aF(U))\nabla^a
\tr\left({\M}U^\dagger+{\M}^\dagger U\right)\right>.
\eeq
The left derivative $\nabla^a$ on $U(N_f)$ is defined through the
relations $\nabla^aU=it^aU$ and $\nabla^aU^\dag=-iU^\dag t^a$ where
$t^a$ are the generators of $U(N_f)$. A simple consistency check on
these is that $\nabla^aUU^\dag=0$. An explicit representation of
$\nabla^a$
is
\beq
\nabla^a & = & i(t^aU)_{ij}\dif{U_{ij}}.
\eeq
We normalize the generators of the unitary group according to
\beq
t^a_{ij}t^a_{kl} & = & \frac{1}{2}\delta_{il}\delta_{jk}.
\label{normgen}
\eeq

Getting a useful Schwinger-Dyson equation is now a matter of choosing
the function $F(U)$ in a clever way. It turns out that with the choice
of $F(U)=U{\M}^\dag$ we arrive at the equation considered in \cite{BN81}
\beq
N_f\left<\tr {\M}^\dag U\right> =
\frac{1}{2}\left< \tr {\M}{\M}^\dag
- \tr U{\M}^\dag U{\M}^\dag\right>,
\label{sdb2matrix1}
\eeq
where we have used (\ref{normgen}).

Now we follow the idea of ref. \cite{BN81}. We know that
\beq
\left<\tr {\M}^\dag U\right> & = & 2\left<\
{\M}^\dag_{ij} \dif{{\M}^\dag_{ij}}\right>\equiv
2\left< D_1\right>, \\
\left<\tr U{\M}^\dag U{\M}^\dag\right> & = &
4\left< {\M}^\dag_{ij}{\M}^\dag_{kl}
\dif{{\M}^\dag_{kj}}\dif{{\M}^\dag_{il}}\right> \nn \\
& \equiv & 4\left< D_2\right>.
\label{sdb2matrix2}
\eeq
To derive how the differential operators $D_1$ and $D_2$ act on the
partition function in terms of the eigenvalues
we use the fact that $\z{2}$ only depends on the $N_f$ functions
\beq
\phi_p & \equiv & \tr (({\M}{\M}^\dag)^p) \nn \\
& = & \sum_i x_i^p,
\eeq
$p=1,\dots,N_f$.
This follows from the fact that $\z{2}$ depends only on the
eigenvalues as well as on permutation symmetry among these.
Using the chain rule we find
\beq
D_1\z{2} & = & \sum_p \frac{\del \z{2}}{\del\phi_p}D_1\phi_p,
\eeq
so we only need to know how $D_1$ operates on the basis functions, i.e.
\beq
D_1\phi_p & = & p{\M}^\dag_{ij}(({\M} {\M}^\dag)^{p-1}{\M})_{ji}\\
& = & p\phi_p.
\eeq
Thus we see that
\beq
D_1 & = & x_i\dif{x_i}.
\eeq
In the same way
\beq
D_2\phi_p & = & \sum_{p'=1}^pp\phi_{p'}\phi_{p-p'} \nn \\
& = & p(p-1)\sum_ix_i^p \nn \\
&& +\sum_{\stackrel{i,j}{i\neq j}}\frac{x_ix_j}{x_i-x_j}
(px_i^{p-1}+px_j^{p-1}),
\eeq
and following \cite{BN81} we see that the second derivative is equivalent
to
\beq
D_2 & = & x_i^2\difto{x_i^2}+\sum_{\stackrel{i,j}{i\neq j}}
\frac{x_ix_j}{x_i-x_j}\left(\dif{x_i}-\dif{x_j}\right).
\label{d2b2}
\eeq
As noted in \cite{BN81} this procedure needs further justification.
From the invariance properties of the partition function we know that
$\z{2}$ is a function of the basis functions only. Accordingly we
should be careful when concluding which form the double differential
operator takes in terms of the eigenvalues.  Expanding the partition
function in terms of the basis functions we see that in principle
things could become complicated from the chain rule when $D_2$ acts on
a product of the basis functions. We can check that this is not the
case by letting our ``guess'' (\ref{d2b2}) act on a product
$\phi_p\phi_q$ and comparing this with the result obtained by letting
the known correct matrix form of the derivative (\ref{sdb2matrix2})
act on $\phi_p\phi_q$. Since we are dealing with a double derivative
no new complications arise if we look at a term
$\phi_p\phi_q\phi_r\dots$ in the expansion of the partition function.
This test, in fact, gives the correct result.

Using (\ref{sdb2matrix1}) we collect
the terms and find the following Schwinger-Dyson equation for
QCD with fermions in a complex representation:
\beq
&& \left[x_i^2\del_i^2+N_fx_i\del_i+
\sum_{\stackrel{i,j}{i\neq j}}\frac{x_ix_j}{x_i-x_j}
\left(\del_i-\del_j\right)\right]\z{2} \nn \\ & = &
\frac{1}{4}\sum_ix_i\z{2}.
\label{sdb2}
\eeq

\subsection{$\beta=1$}
In the case of fermions in a pseudo-real
representation the Goldstone bosons live in the coset $SU(2N_f)/Sp(2N_f)$.
This coset can be parametrized by $UIU^t$, where $I$ is the antisymmetric
$2N_f\times 2N_f$ unit matrix
\beqa
I & = & \left[
\begin{array}{cc}
  0 & \mb{1} \\
  -\mb{1} & 0
\end{array}
\right],
\eeqa
and $U\in U(2N_f)$.
The partition function with vanishing topological charge is \cite{DV01}
\beq
\z{1} & = & \int_{U\in U(2N_f)}\hspace{-7mm}dU e^{\frac{1}{4}\tr\left(
\tilde{\M}^\dag
UIU^t+\tilde{\M}(UIU^t)^\dag\right)},
\eeq
where the mass matrix $\tilde{\M}$
is an arbitrary complex antisymmetric matrix.
Note also that the combination $UIU^t\equiv V$ is antisymmetric, which
will
turn out to
be very useful in the following. Again, the partition function is a
function
of the eigenvalues of $\tilde{\M}\tilde{\M}^\dag$
and since the
Goldstone modes are parameterized by the unitary group we can follow the
same approach as in the case of $\beta=2$. We know that
\beq
0 = \int_{U\in U(2N_f)} \hspace{-8mm}
dU \tr t^a\nabla^a\!\!\left(F(U) e^{\frac{1}{4}
\tr\left(\tilde{\M}^\dag V+\tilde{\M}V^\dag\right)}\right),
\eeq
and if we, inspired by the $\beta=2$ calculation,
choose $F(U)=V\tilde{\M}^\dag$ this results in
\beq
\hspace{-4mm}
\left<(2N_f-1)\tr \tilde{\M}^\dag V\right> & = & \frac{1}{2}\left<\tr
\tilde{\M}
\tilde{\M}^\dag \right. \nn \\ && \left.
\hspace{4mm}- \tr V\tilde{\M}^\dag V\tilde{\M}^\dag\right>\!.
\label{sdb1matrix1}
\eeq
It is important to realize that since both $V$ and ${\M}^\dag$ are
antisymmetric, this equation should be treated a bit differently
from the $\beta=2$ case. Being careful, we define the
differential operators
\beq
\left< \tr \tilde{\M}^\dag V\right> & = & 2\left<
\sum_{\stackrel{i,j}{i\neq j}}
\tilde{\M}^\dag_{ij}\dif{\tilde{\M}^\dag_{ij}}\right> \nn \\
& \equiv &
2\left< D_1\right>,
\label{sdb1matrix2a}\\
\left<\tr V\tilde{\M}^\dag V\tilde{\M}^\dag\right> & = & 4\left<
\sum_{\stackrel{i,j,k,l}{i\neq l, j\neq k}} \tilde{\M}^\dag_{ij}
\tilde{\M}^\dag_{kl}
\dif{\tilde{\M}^\dag_{kj}}\dif{\tilde{\M}^\dag_{il}}\right> \nn \\
& \equiv & 4\left< D_2\right>.
\label{sdb1matrix2}
\eeq
Again we let these act on the basis functions $\phi_p, p=1,\dots,N_f$. The
basis functions
are defined in the same way as in the $\beta=2$ case but differ, since
the matrices in this case are $2N_f\times 2N_f$.
Asuming that the theory has no di-quark terms, the mass matrix has the
form
\beq
\tilde{\cal M} = \left[
\begin{array}{cc}
  0 & {\cal M} \\
  -{\cal M}^t & 0
\end{array}
\right],
\eeq
which results in the basis functions
(again letting $x_i$, $i=1,\dots,N_f$ be the eigenvalues
of the matrix $\M\M^\dag$)
\beq
\phi_p & = & \tr ((\tilde{\M}\tilde{\M}^\dag)^{p}) \nn \\
& = & 2\sum_ix_i^p.
\eeq
Though the sums in eqs. (\ref{sdb1matrix2a}) and (\ref{sdb1matrix2})
do not run over all indices, when doing the actual derivatives the
results turn out rather nicely.  Indeed for the first derivative we
find
\beq
D_1\phi_p & = & \sum_{\stackrel{i,j}{i\neq j}}
\tilde{\M}^\dag_{ij}\dif{\tilde{\M}^\dag_{ij}}\tr((\tilde{\M}
\tilde{\M}^\dag)^p) \nn \\
& = & p\sum_{\stackrel{i,j}{i\neq
j}}\tilde{\M}^\dag_{ij}\left[((\tilde{\M}
\tilde{\M}^\dag)^{p-1}\tilde{\M})_{ji} \right. \nn \\
&& \hspace{17mm} \left. -((\tilde{\M}\tilde{\M}^\dag)^{p-1}
\tilde{\M})_{ij}\right] \nn \\
& = & 2p\phi_p,
\eeq
from which we recognize
\beq
D_1 & = & 2x_i\dif{x_i}.
\eeq
Turning to the second order differential operator
\beq
D_2\phi_p & = & \sum_{\stackrel{i,j,k,l}{i\neq l,
j\neq k}} \!\!
\tilde{\M}^\dag_{ij}\tilde{\M}^\dag_{kl}\dif{\tilde{\M}^\dag_{kj}}
\dif{\tilde{\M}^\dag_{il}}\tr((\tilde{\M}\tilde{\M}^\dag)^p) \nn \\
& = & 2p\sum_{p'=1}^p\phi_{p'}\phi_{p-p'}-2p(p-1)\phi_p \nn \\
& = & 8p\sum_{\stackrel{i,j}{i\neq j}}\frac{x_ix_j}{x_i-x_j}
(x_i^{p-1}-x_j^{p-1}) \nn \\
&& +8p(p-1)\sum_ix_i^p-4p(p-1)\sum_ix_i^p,
\eeq
from which we conclude that
\beq
D_2 = 2x_i^2\difto{x_i^2}+4\sum_{\stackrel{i,j}{i\neq j}}
\frac{x_ix_j}{x_i-x_j}\left(\dif{x_i}-\dif{x_j}\right).
\label{d2b1}
\eeq
Again we have to verify this double derivative on a product
$\phi_p\phi_q$.
The test produces exactly the necessary result.

Collecting the terms using eq. (\ref{sdb1matrix1}) we find the
differential equation
\beq
&\left[x_i^2\del_i^2+(2N_f-1)x_i\del_i+
2\sum_{\stackrel{i,j}{i\neq j}}\frac{x_ix_j}{x_i-x_j}
\left(\del_i-\del_j\right)\right]\!\!\z{1}& \nn \\ & \hspace{-59mm}
= \frac{1}{4}\sum_ix_i\z{1}.&
\label{sdb1}
\eeq

\subsection{$\beta=4$}
For fermions in a real representation the coset is
$SU(N_f)/SO(N_f)$. Once again this can be parameterized by elements
from the unitary group, in this case by $UU^t$. Here the mass
matrix is an arbitrary symmetric complex matrix and the partition
function with vanishing topological charge is \cite{LS92, SV95}
\beq
\z{4} & = & \int_{U\in U(N_f)}\!\!dU e^{\frac{1}{2}\tr\left(
{\M}^\dag UU^t+{\M}(UU^t)^\dag\right)}.
\eeq
Proceeding in the familiar manner, with $W$ being
the symmetric matrix $UU^t$, we find
\beq
0 = \int_{U\in U(N_f)}\hspace{-9mm}dU \tr t^a\nabla^a\left(F(U)
e^{\frac{1}{2}\tr\left({\M}^\dag W+{\M}W^\dag\right)}\right).
\eeq
Defining $F(U)=W{\M}^\dag$ we find the Schwinger-Dyson equation
\beq
\left<(N_f\!+\!1)\!\tr {\M}^\dag W\right> \!=\!
\left<\tr {\M}
{\M}^\dag \!-\! \tr W\!{\M}^\dag W\!{\M}^\dag\right>\!\!.
\label{sdb4matrix1}
\eeq
Having seen the techniques in the anti-symmetric case of $\beta=1$ this
is much the same. The only complication is that in this case the
diagonal elements of the mass matrix are not equal
to zero. Again we carefully define the
differential operators
\beq
\left< \tr {\M}^\dag W\right> & = &
\!\!\left< 2\sum_i{\M}^\dag_{ii}\dif{{\M}^\dag_{ii}}+
\sum_{\stackrel{i,j}{i\neq j}} {\M}^\dag_{ij}\dif{{\M}^\dag_{ij}}\right>
\nn \\
& \equiv &\! \left< D_1\right>, \\
\left<\tr W{\M}^\dag W{\M}^\dag\right>\!\! & = &
\!\left<\sum_{\stackrel{i,j,k,l}{i\neq l, j\neq k}}
\!\!{\M}^\dag_{ij}{\M}^\dag_{kl}\dif{{\M}^\dag_{kj}}\dif{{\M}^\dag_{il}}
\right.\nn \\
&&\! \left.+2\sum_{\stackrel{i,j,k}{j\neq k}}
\!\!{\M}^\dag_{ij}{\M}^\dag_{ki}\dif{{\M}^\dag_{kj}}\dif{{\M}^\dag_{ii}}
\right. \nn \\
&&\! \left.  +2\sum_{\stackrel{i,j,l}{i\neq l}}
{\M}^\dag_{ij}{\M}^\dag_{jl}
\dif{{\M}^\dag_{jj}}\dif{{\M}^\dag_{il}} \right. \nn \\
&&\! + \left. 4\sum_{i,j} {\M}^\dag_{ij}{\M}^\dag_{ji}\dif{{\M}^\dag_{jj}}
\dif{{\M}^\dag_{ii}}\right>\nn \\
& \equiv &\! \left< D_2\right>.
\label{sdb4matrix2}
\eeq
The partition function only depends on the eigenvalues of
${\M}{\M}^\dag$ and since we have returned to working with
$N_f\times N_f$ matrices the $\phi_p$, $p=1, \dots, N_f$
are exactly the same as for $\beta=2$. That is
\beq
\phi_p & = & \tr(({\M}{\M}^\dag)^{p}) = \sum_ix_i^p.
\eeq
In this case we find that
\beq
D_1\phi_p & = & 2p\phi_p,
\eeq
and thus
\beq
D_1 & = & 2x_i\dif{x_i}.
\eeq
The second derivative looks complicated but after some algebra we
end up with the nice expression
\beq
D_2\phi_p & = & 2p\sum_{p'=1}^p\phi_{p'}\phi_{p-p'}+2p(p-1)\phi_p,
\eeq
from which we deduce that
\beq
D_2 = 2\sum_{\stackrel{i,j}{i\neq j}}
\frac{x_ix_j}{x_i-x_j}\left(\dif{x_i}-\dif{x_j}\right)+4x_i^2\difto{x_i^2}.
\eeq
Now we are able to collect the terms and we end up with the result
\beq
&\left[x_i^2\del_i^2+\frac{N_f+1}{2}x_i\del_i+ \frac{1}{2}
\sum_{\stackrel{i,j}{i\neq
j}}\frac{x_ix_j}{x_i-x_j}\left(\del_i-\del_j\right)
\right]\z{4} & \nn \\
& \hspace{-55mm} = \frac{1}{4}\sum_ix_i\z{4}.&
\eeq

\subsection{Including the topological charge}
Including the topological charge in the Schwinger-Dyson equations is now
quite simple. In all classes of chiral symmetry breaking the topological
charge enters the partition function as a change of measure. Again we make
use of the fact that the integral of a total derivative vanishes. Thus we
get an extra term in the Schwinger-Dyson equations corresponding to the
derivative of the change in measure.

More specifically, consider first $\beta=2$. Allowing a non-vanishing
topological charge changes the measure in (\ref{partapp2}) by
$dU\to dU(\det U)^\nu$. Thus we need to know the derivative of $\det U$
\beq
\nabla^a(\det U) & = & i(t^aU)_{ij}\dif{U_{ij}} \det U \nn \\
& = & i(t^aU)_{ij}U^{-1}_{ji}\det U \nn \\
& = & i\tr (t^a) \det U
\eeq
since
\beq
\det(U+\delta U) & = & \det(U(1+U^{-1}\delta U)) \nn \\
& = & \det U(1+\tr(U^{-1}\delta U)\!+\!\dots)\!.
\eeq
Thus we see that the expectation value equation (\ref{sdb2matrix1})
changes into
\beq
\hspace{-2mm}(N_f\!+\!\nu)\!\left<\tr {\M}^\dag U\right>\! = \!
\frac{1}{2}\!\left< \tr {\M}{\M}^\dag
\!-\! \!\tr U{\M}^\dag U{\M}^\dag\right>\!,
\eeq
which only alters the coefficient in front of the differential operator
$D_1$.
Modifying the partition function, of course also implies that the
dependencies
are changed. As discussed in section \ref{vcsec}, $(\det
\M)^{-\nu}\z{2}_\nu$
is a
function of the eigenvalues of $\M\M^\dagger$ only, and all results of the
last
section are thus valid in sectors of non-vanishing topological charge
provided
we change the coefficient in front of $D_1$ and let the differential
operators
act on $(\det \M)^{-\nu}\z{2}_\nu$ instead of just $\z{2}$. Thus we find
\beq
&& \left[x_i^2\del_i^2+(N_f+\nu)x_i\del_i+
\sum_{\stackrel{i,j}{i\neq j}}\frac{x_ix_j}{x_i-x_j}
\left(\del_i-\del_j\right)\right] \nn
\\ & \times & (\det \M)^{-\nu}\z{2}_\nu =
\frac{1}{4}\sum_ix_i(\det \M)^{-\nu}\z{2}_\nu.
\label{sdeqapp2}
\eeq

Turning to $\beta=1$ where the measure also changes by $dU\to dU (\det
U)^\nu$
we find that the expectation value equation (\ref{sdb1matrix1}) changes
into
\beq
\hspace{-4mm}
& \left<(2N_f-1+\nu)\tr \tilde{\M}^\dag V\right> & \nn \\
= & \frac{1}{2}\left<\tr
\tilde{\M}
\tilde{\M}^\dag - \tr V\tilde{\M}^\dag V\tilde{\M}^\dag\right>. &
\eeq
In this case we need to replace $\z{1}$ with
$(\det\tilde{\M})^{-\nu/2}\z{1}_\nu=(\det\M)^{-\nu}\z{1}_\nu$ and we find
the
differential equation
\beq
& \left[x_i^2\del_i^2+(2N_f-1+\nu)x_i\del_i+
2\sum_{\stackrel{i,j}{i\neq j}}\frac{x_ix_j}{x_i-x_j}
\left(\del_i-\del_j\right)\right]& \nn \\
&\hspace{-5mm}\times (\det\M)^{-\nu}\z{1}_\nu
\, = \, \frac{1}{4}\sum_ix_i(\det\M)^{-\nu}\z{1}_\nu.&
\label{sdeqapp1}
\eeq

Finally, for $\beta=4$ the change of the measure is
$dU\to dU (\det U)^{2\bar{\nu}}$, with $\bar{\nu}=N_c\nu$. Hence eq.
(\ref{sdb4matrix1}) becomes
\beq
&\left<\left(\frac{N_f+1}{2}+\bar{\nu}\right)\tr {\M}^\dag W\right> & \nn
\\
=&\frac{1}{2}\left<\tr {\M}
{\M}^\dag - \tr W{\M}^\dag W{\M}^\dag\right>.&
\eeq
$\z{4}$ should be replaced by $(\det \M)^{-\bar{\nu}}\z{4}_{\bar\nu}$ and
thus
\beq
&\left[x_i^2\del_i^2+\left(\frac{N_f+1}{2}+\bar{\nu}\right)x_i\del_i+
\frac{1}{2}
\sum_{\stackrel{i,j}{i\neq
j}}\frac{x_ix_j}{x_i-x_j}\left(\del_i-\del_j\right)
\right]& \nn \\
& \! \times (\det \M)^{-\bar{\nu}}\z{4}_{\bar\nu} \, = \,
\frac{1}{4}\sum_ix_i(\det \M)^{-\bar{\nu}}\z{4}_{\bar\nu}\!.&
\label{sdeqapp4}
\eeq


\section{The Topological Susceptibility \label{topsusapp}}
In this appendix we calculate the topological susceptibility. This we
calculate by allowing a non-vanishing vacuum angle $\theta$ and
subsequently projecting onto sectors of fixed topological charge by
means of a Fourier transform. We consider the classes of symmetry
breaking separately.

The case of $\beta=2$ has been considered previously \cite{D01} by means
of the same method as we will use here. The result found was
\beq
\left<\nu^2\right>_{\beta=2} & = & \frac{m_0^2VF^2}{2N_c}.
\label{topsus2}
\eeq

Matching of the two finite volume regimes requires that the dependence
of the topological susceptibility upon the Dyson index follows the
same pattern as the last (non-trivial) part of the diagonal
propagator.  Thus we expect the topological susceptibility to be twice as
large as formula (\ref{topsus2}) in the $\beta=1$ case while being
identically the same in the $\beta=4$ case.

\subsection{$\beta=1$}
A non-vanishing vacuum angle can be absorbed into the mass-matrix
$\cal M$ by letting $\M\to e^{i\theta/N_f}\M$ \cite{LS92, SV95}.  We
can include the vacuum angle in a shifted $\Xi_0$. To see this
consider how the change of the mass matrix affects the combination
\beq
\mbox{Re}\tr\M U^\dag \! \to \! \mbox{Re}\tr\left[\M e^{i\theta/N_f}
e^{-i\sqrt{2}\xi/F_1}e^{-i\sqrt{2}\varphi_0/F_1}\right]\!\!,
\eeq
from which we find that we can include the vacuum angle in a change of
variables
\beq
\varphi_0 & \to & \varphi_0-\frac{\theta F_1}{\sqrt{2}N_f}.
\eeq
Taking the trace of this equation we obtain (note that the dimension
of the coset is $2N_f\times 2N_f$)
\beq
\Xi_0 \to  \Xi_0' \equiv \Xi_0-\sqrt{2}\theta F_1.
\eeq

Finally we perform the Fourier transform (for large volumes)
\begin{widetext}
\beq
\z{1}_\nu & = & \frac{1}{2\pi}\int_{-\pi}^{\pi}d\theta e^{-i\nu\theta}
\int_{SU(2N_f)/Sp(2N_f)}dU_0 \exp\left[\frac{1}{2}\tr\M(U_0+U_0^\dag)
-\frac{m_0^2V}{2N_c}\Xi_0'^2\right] \nn \\
& \propto & \sqrt{\frac{N_c}{2m_0^2VF_1^2}}e^{-\nu^2N_c/4m_0^2VF_1^2}
\int_{U(2N_f)/Sp(2N_f)}dU_0 (\det U_0)^{-\nu/2}
\exp\left[\frac{1}{2}\tr\M(U_0+U_0^\dag)\right],
\eeq
\end{widetext}
from which we identify the topological susceptibility
\beq
\left<\nu^2\right> & = & \frac{2m_0^2VF_1^2}{N_c} = \frac{m_0^2VF^2}{N_c}.
\eeq
As expected, this is exactly twice the result from the $\beta=2$ case.

\subsection{$\beta=4$}
The $\beta=4$ calculation proceeds in the same way. In this case the
vacuum angle can be included in the mass matrix by letting $\M\to
e^{i\theta/N_cN_f}\M$ \cite{LS92, SV95}. Again this can be included in the
field of constant modes, this time by the substitution
\beq
\varphi_0 & \to & \varphi_0-\frac{\theta F_4}{\sqrt{2}N_cN_f}.
\eeq
Thus for the traced equation we find
\beq
\Xi_0 \to \Xi_0' \equiv \Xi_0-\frac{\theta F_4}{\sqrt{2}N_c}.
\eeq
The projection onto sectors of fixed topological charge is performed as
before
\begin{widetext}
\beq
\z{4}_\nu & = & \frac{1}{2\pi}\int_{-\pi}^{\pi}d\theta e^{-i\nu\theta}
\int_{SU(N_f)/SO(N_f)}dU_0 \exp\left[\frac{1}{2}\tr\M(U_0+U_0^\dag)
-\frac{m_0^2V}{2N_c}\Xi_0'^2\right] \nn \\
& \propto & \sqrt{\frac{1}{\left<\bar\nu^2\right>}}e^{-\bar\nu^2/2\left<
\bar\nu^2\right>}
\int_{U(N_f)/SO(N_f)} dU_0 (\det U_0)^{-\bar\nu}
\exp\left[\frac{1}{2}\tr\M(U_0+U_0^\dag)\right],
\eeq
\end{widetext}
with the topological susceptibility given by
\beq
\left<\bar\nu^2\right> & = & \frac{m_0^2VF_4^2}{2N_c} =
\frac{m_0^2VF^2}{2N_c}.
\eeq
Again we define $\bar\nu\equiv N_c\nu$. This result is also as expected.


\end{document}